\documentclass[%
 reprint,
 amsmath,amssymb,
 aps,
]{revtex4-2}

 \usepackage{flushend}

\usepackage{graphicx}
\usepackage{dcolumn}
\usepackage{bm}
\usepackage{graphicx}
\usepackage{xcolor}
\usepackage{float}

\begin{document}

\preprint{APS/123-QED}

\title{Description of mesoscale pattern formation in shallow convective cloud fields by
	using time-dependent Ginzburg-Landau and Swift-Hohenberg stochastic equations}

\author{Diana L. Monroy}
\author{Gerardo G. Naumis}%
\email{e-mail: naumis@fisica.unam.mx}
\affiliation{%
	Departamento de Sistemas Complejos, Instituto de F\'isica, \\ Universidad Nacional Aut\'onoma de M\'exico\\
	Apdo. Postal 20-364, 01000, Ciudad de M\'exico, CDMX, MEXICO.
}%


\begin{abstract}
	The time-dependent Ginzburg-Landau equation and the Swift-Hohenberg equation, both added with a stochastic term, are proposed to describe cloud pattern formation and cloud regime phase transitions of shallow convective clouds organized in mesoscale systems. The starting point is the Stechmann-Hottovy linear spatio-temporal stochastic model for tropical precipitation, used to describe the dynamics of water vapor and tropical convection. By taking into account that shallow stratiform clouds are close to a self-organized criticallity and that water vapor content is the order parameter, it is observed that sources must have  non-linear terms
	in the equation to include the dynamical feedback due to precipitation and evaporation. The non-linear terms are derived by using the known mean field of the Ising model, as the  Stechmann-Hottovy  linear model presents the same probability distribution. The inclusion of this non-linearity leads to a kind of time-dependent Ginzburg-Landau stochastic equation, originally used to describe superconductivity phases. By performing numerical simulations,  pattern formation is observed. These patterns are better compared with real satellite observations than the pure linear model. This is done by comparing the spatial Fourier transform of real and numerical cloud fields. However,  for highly ordered cellular convective phases, considered as a form of Rayleigh-B\'enard convection in moist atmospheric  air, the Ginzburg-Landau model does not allow to reproduce such patterns. Therefore, a change in the form of the small-scale flux convergence term in the budget moist atmospheric air is proposed. This allows to derive a Swift-Hohenberg equation. In the case of closed cellular and roll convection, the resulting patterns are much more organized that the ones obtained from the Ginzburg-Landau equation  and better reproduce satellite observations, as for example, horizontal convective rolls fields.
	
\end{abstract}

\maketitle

\par
\section{Introduction} Convective clouds are well known to be crucial components of weather and climate, being a key process not only in the transport of heat, moisture, momentum, and dynamical quantities in the atmosphere but also by strongly affecting solar and long-wave radiation budgets from local to global scales \cite{Schneider, Nuijens}. Historically, most research involving convective clouds has focused on deep rather than shallow clouds. However, shallow convective clouds have significant impacts on the mesoscale as well as for large scale atmospheric dynamics \cite{Deng}.\par
The study of shallow clouds is worthy for at least two reasons: first, they cool our planet reflecting a significant portion of the incoming solar radiation back to space contributing only marginally to the greenhouse effect; and second, shallow clouds cover large fractions of our planet’s sub-tropical oceans \cite{Rasp, Nuijens}. Even changes in the order of $1 \%$ in cloud cover or other properties may significantly affect the overall radiation balance \cite{Dagan}. As a consequence, cloud feedback influences significantly the response of the climate system to global warming \cite{Schneider, Ceppi}. \par
Shallow clouds exhibit spatial organization over a wide range of scales  \cite{Nuijens, Khouider}. Compared to spatially homogeneous low clouds, these modes of organization could be significant for the radiative effect of  convective organization. They presumably affect the interaction of convection with atmospheric humidity and thus cloudiness plays a role in climate variability \cite{Tobin}. Cloud systems formed by shallow convection have horizontal dimensions ranging from several to $100$ or $200$ kilometers. They are often characterized as mesoscale patterns \cite{Houze} and are largely ignored in actual climate models \cite{Rasp}. \par
Therefore, mesoscale systems need to be considered in climate-model parameterizations of the physical processes that affect the shallow clouds radiative response to climate perturbations \cite{Vogel}. At the same time, this is one of the challenges in climate sciences as contemporary climate models cannot resolve the length scales where it occurs \cite{Nuijens}. Even the driving mechanisms responsible for these patterns are not completely well understood \cite{Taylor}. \par
Stratocumulus clouds (Sc) are relevant examples of mesoscale organization of shallow convection on stratiform cloudiness. They have been studied in recent years due to their impact on the amount of sunlight reflected back to space \cite{Colleen, Schneider}. Covering approximately one-fifth of Earth’s surface in the annual mean, Sc are the dominant cloud type by area covered. Thus, there are few regions of the planet where these clouds are not climatologically important \cite{Wood}. Sc are characterized by honeycomb-like patterns of stratiform cloudiness, arranged in either ‘open’ or ‘closed’ cells controlled by processes from the micrometer to the kilometer scale which interact in and above the scale O(10-100km) of large-scale models \cite{Noteboom}.  \par
The organization of Sc into cellular or roll convection could be considered in first approximation as a form of Rayleigh–Bénard convection in the atmospheric boundary layer \cite{Agee}. However, this mechanism does not completely explain the multiscale turbulent character of the mesoscale cloud convection (MCC) seen in observations, whereby other theories have been proposed to explain the driving of these patterns \cite{Blossey}.  For Sc, in addition to the temperature difference between the lower boundary (the sea or land surface) and the upper boundary (a subsidence inversion), there are extra factors and processes whose interaction results in an enhancement or damping of the atmospheric convective circulation \cite{Noteboom}. \par

\begin{figure}[t]
	\centering
	\includegraphics[width=1.0\linewidth]{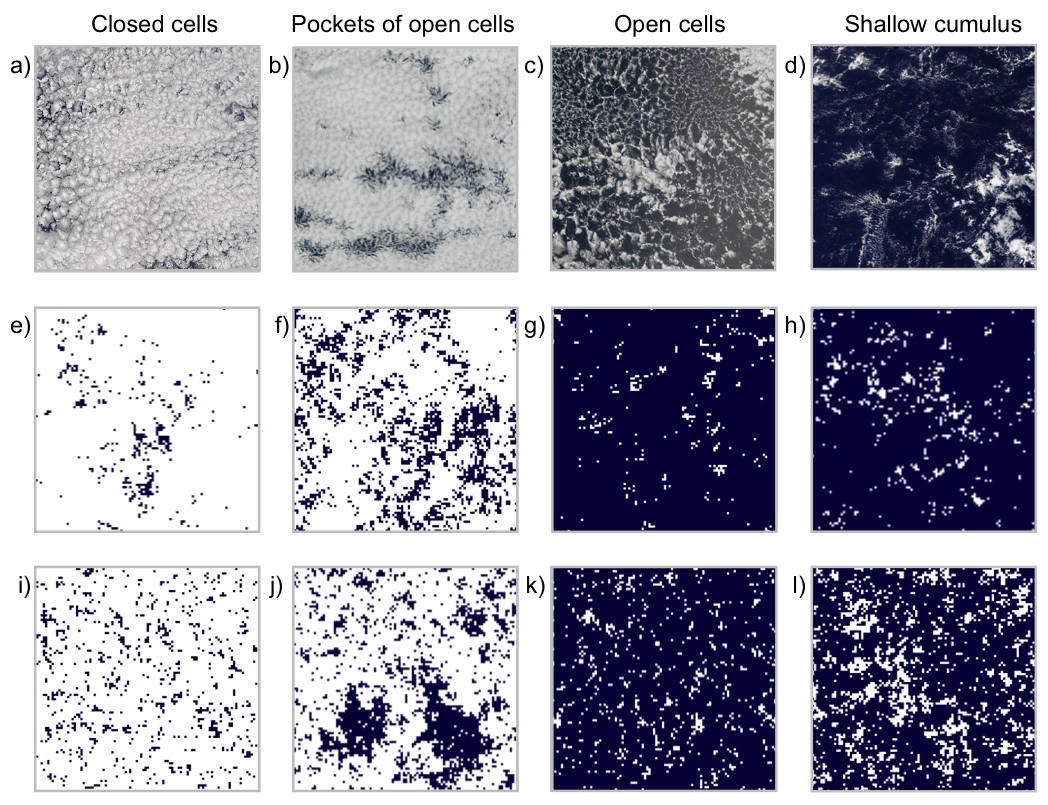}
	\caption{The four distinctive phases of shallow cloud organization: closed-cell stratocumulus,  pockets of open-cell stratocumulus,  open-cell stratocumulus, and shallow cumulus viewed from satellite in panels a) to d), generated by the HS model (Eq. \ref{SHmodel}) with the parameters proposed in Ref \cite{Hottovy} in  panels e) to h) and by the non-linear idealized model (Eq. \ref{Ginzburg})  in panels i) to l). See Supplemental Material for the parameter values \cite{Supplemental}.  The data of the real fields was taken from the Moderate Resolution Imaging Spectroradiometer (MODIS) data, and from the Geostationary Satellite Server (GOES) data from NOAA.}
	\label{Phases}
\end{figure}
Many of those processes are key in stratocumulus and MCC clouds: short-wave heating and long-wave cooling at cloud top, turbulence and entrainment , precipitation, latent heating, evaporative cooling and surface fluxes of energy as well as microphysical processes closely related with droplets concentration, aerosol effect and their influence in drizzle formation \cite{Wood}. It is important to note the different processes involved in each regime. While open cells (Fig. \ref{Phases}c) appear as a consequence of descending motion and sinks of clear air at centers with ascending and cloudy air at their borders, closed cells (Fig. \ref{Phases}a) are formed in presence of upward motion and cloudy air in their centers and descending air at their interfaces. Heating from below is the key responsible process in open-cell convection when there is a large difference between sea surface temperature and air temperature; instead of that, radiative cooling of cloud tops is the key responsible process for closed-cell convection \cite{Wood, McCoy, Noteboom}.\par
The transition from closed to open cellular convection is interesting from the system dynamics as well as from the perspective of radiative forcing of the climate but is not clearly understood yet.  Many theoretical and numerical models have been proposed. Two of the most investigated mechanisms are (1) cloud‐aerosol‐precipitation interactions \cite{Koren} and (2) advection over warmer water \cite{Yamaguchi, Nesbitt, Feingold}. The first approach can be thought of as microphysically driven and the second one as large‐scale meteorologically driven. This last mechanism has been studied in recent years using satellite data, proposing a relationship between column-integrated water and precipitation rate as a Self-Organized Criticality (SOC) \cite{Bak} system. According to this, a critical value of water vapor (the tuning parameter) determines a non-equilibrium continuous phase transition to a regime of strong atmospheric convection with the emergence of precipitation (the order parameter)\cite{Peters}.\par
Based on this ideas, Hottovy and Stechmann proposed a linear stochastic equation to describe cloud phase transitions \cite{Stechmann}. In this paper, we propose to modify such model by including  a feedback mechanism for sources and sinks like precipitation or evaporation. This leads to a time-dependent stochastic Ginzburg-Landau equation and if convection is included, to a time-dependent stochastic Swift-Hohenberg equation. Such equations describe the formation and transition of stratocumulus cloud regimes: open cells, closed cells, and pockets of open cells \cite{Stevens} (Fig. \ref{Phases}b), as well as an unrobust phase (Fig. \ref{Phases}d) observed in shallow clouds. This mechanism for organized mesoscale convection simulates the transition to strong convection as a result of an increase in precipitation rate as a function of the column water vapor (CWV), in particular, for stratiform rain systems as Sc clouds \cite{Ahmed}. By means of Fourier transforms, we compare the obtained patterns with several real cloud fields obtaining a good agreement.\par
 In fact, the idea of developing a Ginzburg-Landau-type equation for cloud patterns is not completely new. In 2013, Craig and Mack proposed a Cahn-Hilliard equation to build a coarsening model for self‐organization of tropical convection \cite{Craig_2013}. Their model started with the Allen-Cahn equation, which generalizes the Ginzburg-Landau  equation to more general functionals \cite{Allen_1979}. As in our work, they used a similar order parameter, the tropospheric humidity, and a budget equation with feedback. They found a phase transition
when  the Landau-type functional has two minima, rather than one, leading to a bistable system with two equilibrium values of humidity \cite{Craig_2013}. Beyond the not so important differences in the type of Landau functional, the main departure from our work is that here we include stochastic terms in the equations. Thus, noise is considered in the time evolution, while in the work by Craig and Mack the noise is only used to produce an initial state \cite{Craig_2013}. As in other systems, noise has important effects in the pattern formation phase diagram \cite{Sancho_1999,Leocher_2003}.\par
The structure of this paper is the following, in section \ref{sec:Linear} we detail the linear model while in sections \ref{sec:Ginzburg} and \ref{sec:Swift} the non-linear models are introduced. Finally, the conclusions are given in section \ref{sec:conclusions}.

\section{The Hottovy and Stechmann linear Stochastic Model for mesoscale shallow patterns} \label{sec:Linear}

In this section, we explain the basic details of the Hottovy and Stechmann  (HS) model \cite{Stechmann}, based upon a idealization of water vapor dynamics as a stochastic diffusion process. In this model, several effects of the physical processes involved in cellular convection are included: evaporation, turbulent advection–diffusion of water vapor and precipitation.\par
The HS Model \cite{Stechmann} was proposed as a model for the dynamics of the cloudy boundary layer following the idealized simplification of models of phase transitions in other contexts. The model starts by considering the evolution of the  total moisture content  $q=q(\mathbf{r} ,t)$ (water vapor plus condensed water, $i.e$, liquid and ice) in each planetary boundary layer (PBL) column at a horizontal spatial location $(x, y)$, normalized and shifted so that $q = 0$ represents the saturation level \cite{Hottovy}.
Spatio-temporal changes, given by the convective derivative of $q$, must be equal to the contribution of all sources and sinks such as precipitation or evaporation,
\begin{equation}
\frac{Dq}{Dt}=\frac{\partial q}{\partial t}+\mathbf{v} \cdot \mathbf{\nabla} {q}=S
\label{eq:balance}
\end{equation}
where $\mathbf{v}$ is the velocity. We next decompose $q$ as  $q=\bar{q}+q'$, where $\bar{q}$ is a large-scale average component and $q'$ is a small fluctuation part, and in a similar way we decompose $\mathbf{v}=\bar{\mathbf{v}}+\mathbf{v}'$. Using Eq. (\ref{eq:balance}), we obtain an equation for the large component \cite{Hottovy},
\begin{equation} \label{eq:budget}
\frac{\partial \bar{q}}{\partial t}=\bar{S}-\mathbf{\nabla}\cdot \left( \bar{q}\bar{\mathbf{v}}\right)-\mathbf{\nabla} \cdot \left( \overline{q'\mathbf{v'}}\right)
\end{equation}
where it was used that $\bar{q'}=0$ and  $\overline{v'}_x=\overline{v'}_y=0,$. Next the small-scale ﬂux convergence term  $\mathbf{\nabla}\cdot \left( \overline{q'\mathbf{v'}}\right)$ is approximated by a laplacian $b \nabla^{2}q$, used to represent eddy diffusion and mixing due to turbulence. The parameter $b$ is an effective diffusion constant. The nonlinear turbulent effects contained in $\mathbf{\nabla} \left( \bar{q}\bar{\mathbf{v}}\right)$ are taken into account by additional turbulent damping \cite{Betts} $-\overline{q}/\tau_0$ and stochastic forcing, $D \dot {W}$ \cite{Majda}. The term $\overline{q}/\tau_0$ represents a relaxation, where the parameter $\tau_0$ is obtained through a careful analysis of the column-integrated water and precipitation rate \cite{Hottovy}. The term $D\dot{W}$ represents a stochastic forcing, and is used as the simplest model for the turbulent fluctuations and others physical processes with a random component, such as the entrainment. Finally, the source term $\bar{S}$ represents the net water sources and sinks, including precipitation and evaporation of water from the ocean surface. It is considered to contribute with a constant and deterministic forcing $F_0$, and a partial stochastic contribution, taken already into account in the constant $D$. \par
Finally, the temporal evolution is given by the following equation\cite{Hottovy},
\begin{equation}
\frac {\partial q}{\partial t}  =  b\nabla ^{2}q  - \frac{1}{\tau_0} q+ F_0 + D\dot{W} 
\label{SHmodel}
\end{equation}
where here, and to avoid overburden the notation, $q$ represents the average part $\overline{q}$. In what follows, the same convention will be used. \par
It has been shown that this model can be translated into a spin-like Hamiltonian system which presents phase transitions\cite{Stechmann} once $q$ discretized using a function that takes the values $0$ or $1$ depending on the sign of $q$. Typical clouds fields obtained through numerical simulations using this equation are shown in Fig. \ref{Phases}. Therein, we include real images from satellite to provide a comparison. \par
\begin{figure}[t!]    \centering
	\includegraphics[width=1.0\linewidth]{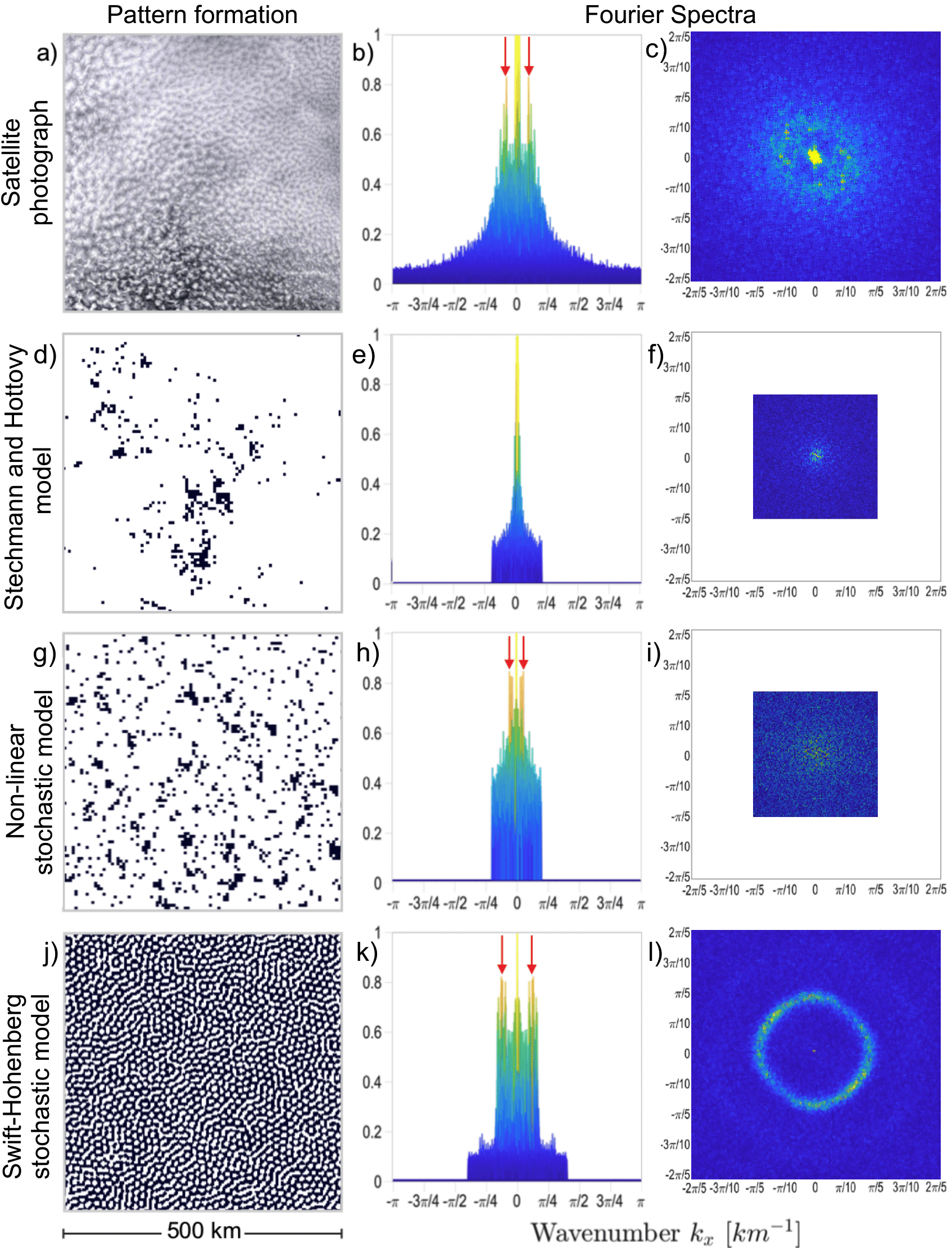}
	\caption{Fourier transform of the closed-cell phase. Panels in the left column show the cellular pattern taken from a) satellite photograph, d) Hottovy and Stechmann model, g) Ginzburg-Landau stochastic model and j) Swift-Hohenberg stochastic model.  In the central and right column we present the Fourier spectra of each pattern in the $I/I_0-k_y$ plane and in the orthogonal plane, respectively. We can identity in panels b-c) and k-l) a dominant frequency with radial symmetry indicated by red arrows, corresponding to a characteristic length of $ \approx 14$km.
		The maximal spatial frequencies in panels  e), h) and k) are determined by the resolution of the grid used in the simulation given in the units of $k_x$ (see text). See Supplemental Material for the parameter values \cite{Supplemental}.  The data of the real fields was taken from the Moderate Resolution Imaging Spectroradiometer (MODIS) data and from the Geostationary Satellite Server (GOES) data from NOAA.}
	\label{Fig:FourierAndFields}
\end{figure}
Although the model is able to reproduce the overall aspect of the fields and the phase transitions between them, it is also clear that there is much more organization in real cloud patterns for closed phases. To account for this, we have calculated the spatial Fourier transform of real closed-cell patterns taken from satellite photographs as well as from the outcome of HS model, as seen in Fig. \ref{Fig:FourierAndFields}. \par
In Fig. \ref{Fig:FourierAndFields} panels b) and c) we can identify one spatial frequency (wave-vector) that reveals the existence of a particular structure. This is very clear in
\ref{Fig:FourierAndFields} c), in which a ring-like structure is observed. Nevertheless, in  Fig. \ref{Fig:FourierAndFields}, panels e) and f), we see that the Fourier transform of the outcomes obtained from the HS model does not show any characteristic dominant structure. This is expected as the HS is a linear model which does not couple modes \cite{Stechmann}.\par
Notice that in the case of the satellite photographs, we adjust the contrast and exposure of the original image -showed in Fig. \ref{Fig:FourierAndFields}a)- before converting the grayscale image into a binary image. This is done to define the cells with more details and precision. \par
Also, observe that in Fig. \ref{Fig:FourierAndFields}  e)-f), h)-i) and k)-j) there is a lower cut-off of the spectrum when compared  with Fig. \ref{Fig:FourierAndFields} b)-c) and k)-l). This is due to the resolution of the grid used. Although one can increase the cut-off frequency  by growing the number of points in the simulation mesh, it turns out that the phases and parameters of the HS model depend upon the mesh. On the other hand, decreasing the resolution of the real cloud fields leads to a lower-quality Fourier image. A trade-off is thus needed to keep the original parameters of the  HS model and the best resolution of the real cloud fields. To solve this conundrum, here we adopted the policy of using absolute units in reciprocal space. These units are determined by the length ($L=500$) in Km of the real space field and the resolution of the photograph ($N_{pixels} \times N_{pixels}=500\times 500$), resulting in the cut-off frequency $k_x=\pm \pi N_{pixels} /L=\pm \pi \ [Km^{-1}]$. For the simulation, the mesh has $N \times N$ points resulting in a cut-off frequency $k_x=\pm \pi N /L= \pm \pi (N/500) [Km^{-1}]$.
In all the Fourier transforms, the intensity $I$ is scaled by the maximal intensity $I_0$.

\section{Non-linear model: time-dependent Ginzburg-Landau stochastic equation}
\label{sec:Ginzburg}

One of the most important points in the work of Craig and Mack  and HS is the recognition of $q$ as an order parameter \cite{Craig_2013,Stechmann}. In general, pattern formation is governed by order parameters whose spatio-temporal behavior is determined by nonlinear partial differential equations \cite{Gurevich}.
This suggests that the extra features seen in real cloud patterns are due to non-linear effects. Following this idea, here we consider the cellular convective pattern described by a state vector $p(\mathbf {r} ,t)$ which in this case corresponds to the cloud cover. Its evolution equation takes the general form of a partial differential equation\cite{Gurevich}:

\begin{equation}
\frac {\partial p (\mathbf {r} ,t)}{\partial t}  = \textbf{N} \left[\nabla, p (\mathbf {r},t)\right]
\label{eq:Ndef}
\end{equation}

where $\textbf{N}$ denotes a nonlinear function.
The behavior of the state vector $p(\mathbf {r} ,t)$ of the pattern forming system can be represented as a functional of one or several order parameters, denoted by $\Phi(\mathbf {r} ,t)$ that often can be directly related to a physical observable \cite{Gurevich},

\begin{equation*}
	p (\mathbf {r} ,t) = Q \left[\Phi(\mathbf {r},t)\right]
\end{equation*}

where $Q$ is a functional of $\Phi(\mathbf {r},t)$.
In order to recover the linear equation proposed by HS, in our model we identify $\Phi(\mathbf {r} ,t) = q(\mathbf {r} ,t)$, $i.e.$, the CWV in each column of the lattice. Thus, instead of solving the determining equations for the state vector $p(\mathbf {r} ,t)$, the spatio-temporal evolution is in general determined by an equation for the order parameter field \cite{Gurevich}.
The most simple case is the following,
\begin{equation}
\frac {\partial q}{\partial t}  = L(\Delta) q +  N\left[q\right]
\label{eq:SHNonModel}
\end{equation}

Here $L(\Delta)$ is a linear operator and $N\left[q,t)\right]$ the nonlinear functional that is approximated by a polynomial expansion of $q$ in its low order derivatives.\par
Therefore, by comparing with Eq. (\ref{SHmodel}) we can identify the operator $L(\Delta)$ with $\tau_0^{-1} + b\nabla ^{2}$, while $D$ and  $F_0$ are parameters that determine the strength of the random and deterministic forcing generated by internal forcing due to small scale cloud processes and large-scale external forcing, respectively. The transition of cloud area fraction ($CAF$) from a regime of closed cellular convection to a regime of pockets of open cells is determined by both parameters  \cite{Khouider}.\par
 Let us start with the simple model given by Eq. (\ref{eq:SHNonModel}) to indicate how non-linear terms arise. We start by pointing out that several observational data and numerical  studies have documented  the crucial relationship between precipitation and water vapor  for  precipitation prediction in the context of convective parametrizations. Peters and Neelin \cite{Peters,Nesbitt} showed that there is a critical value $q_c$ of the CWV where the mean precipitation $\langle P(q) \rangle$ increases rapidly as an approximate power law, i.e.,  $\left\langle P(q) \right\rangle  \sim (q - q_c)^{\beta}$, for  $q > q_c$. 
As $\beta < 1$, the precipitation variance has a strong peak at the critical value $q_c$ and then diminishes \cite{Neelin, Holloway, Bretherton}.  \par
It has been argued that the mechanism presents a tendency to self-maintain at criticality instead of being simply controlled by an external parameter \cite{Peters,Nesbitt}. In fact, self-organized critically (SOC) has been proposed to describe  macroscopic critical phenomena such as organized structures associated with atmospheric convection \cite{Yano}.\par
This organization mechanism is supported by observations which exhibit that, even when the system hardly exceeds $q_c$, the CWV tends to decay more slowly than an exponential rate toward the higher values, reflecting the tendency towards SOC \cite{Peters,Noteboom}. The same studies show a scale invariance suggesting a scaling law  for atmospheric convection. Moreover, the invariance under spatial averaging suggests the applicability of the renormalization group (RNG), also supported by the SOC approach \cite{Peters,Noteboom}. \par
In the original HS model, the relaxation time $\tau_0^{-1}$ and the forcing $F_0$ was adjusted in such a way that different assumed models of the precipitation ratio fitted the results of Peters and Neelin for the precipitation conditional probability. 
If $r_{i,j}$ is the precipitation ratio for a cell with integer coordinates $(i,j)$ in a square mesh, there are two precipitation models, the first model is the Betts–Miller-like rain rate model \cite{Betts},
\begin{equation}
r_{i,j}=|F_0|\sigma_{i,j}
\end{equation}
the other was provided by HS \cite{Hottovy},
\begin{equation}
r_{i,j}=\left[ |F_0|+q_{i,j}/\tau_0\right]\sigma_{i,j}
\end{equation}
where $\sigma_{i,j}=1$ if $q>0$, and  $\sigma_{i,j}=0$ otherwise. 
Notice that $\sigma_{i,j}$ is analogous to a spin variable. Its role is to signal whenever $q$ is above the precipitation threshold $q=0$. Then is possible to have rain. \par
While the conditional probability for precipitation can be obtained from the distribution function of $q$, {\it the linear model does not provide a feedback threshold due to precipitation in the source term $\bar{S}$}. In other words, the precipitation can be calculated a posteriori once the model is solved, but it does not enter into the calculation. We require $\overline{S}$ to depend upon $q$.\par
Therefore, to improve the model one needs to include the fact that once the threshold for precipitation is reached, indicated by the spin variable $\sigma_{i,j}$, the source term will change. In fact, $\sigma_{i,j}$ can be used to derive an equivalent Ising Hamiltonian for the cloud field \cite{Hottovy}. Now comes the question, what is the most simple and natural choice for the feedback term? Following the Ising analogy, we can replace the spins $\sigma_{i,j}$ by the known Ising mean field, $\overline \sigma \approx (1+tanh(q/T))/2$ with $T$ a constant. Notice how the field is shifted to have $\sigma_{i,j} \approx \overline{\sigma}=0$ for $q\rightarrow -\infty$ and $\overline{\sigma}=1$ for $q\rightarrow \infty$ . This results on two possible average precipitation rates  $\overline{r}$ depending upon the used model,
\begin{equation}
\overline{r} \approx \frac{1+tanh(q/T)}{2}|F_0| \\
\label{eq:simple1}
\end{equation}
or,
\begin{equation}
\overline{r} \approx \frac{1+tanh(q/T)}{2}\left[ |F_0|+\frac{q}{\tau_0}\right]
\label{eq:precipimodel2}
\end{equation}
As we are interested in the region around the threshold, $i.e.$, near the lineal model, we can expand the hyperbolic tangent  to obtain, using Eq. (\ref{eq:simple1}), 
\begin{equation}
\overline{r} \approx \left(1+ \frac{q}{T}-\frac{1}{3}\left(\frac{q}{T}\right)^{3}+\frac{2}{15}\left(\frac{q}{T}\right)^{5}+...\right)\frac{|F_0|}{2}
\end{equation}
Thus, we generated a non-linear term able to model dynamically a precipitation threshold. Although in principle we can just modify  the sources term in Eq. (\ref{SHmodel}) by using $\bar{S} \rightarrow \bar{S}-\bar{r}$, it will be unwise not to recognize that  sources must also depend dynamically on $q$, as for example, the conditional probability of having an increased $q$ grows once precipitation occurs \cite{Lebsock, HollowayA}. Thus, we left open the possibility of having an interplay between sources and sinks by the replacement $\bar{S} \rightarrow  F_0 +D\dot{W}-\bar{r}+\bar{s}$ where  $\bar{s}$ is an average dynamic source. The most simple model is to assume $\bar{s} \approx f\bar{r}$ where $f$ controls the relative weight between sources, like evaporation, and precipitation. The parameter $f$ allows an interplay between two kinds of non-linear regimes, one dominated by sinks the other by sources.\par
Finally, we include, up to third order, the sources and sinks terms in Eq. (\ref{SHmodel}) to obtain the following non-linear model built from Eq. (\ref{eq:simple1}) Betts–Miller-like rain rate precipitation model, 
\begin{equation}
\begin{split}
\frac {\partial q}{\partial t} & = b\nabla ^{2}q +Eq - Kq^{3} \\ & +  D\dot{W} + F
\label{Ginzburg}
\end{split}
\end{equation}
where the constants are given by, 

\begin{equation}
E =\frac{1}{\tau_s}-\frac{1}{\tau_0}, \hspace{0.2cm}
K=\frac{1}{3\tau_s T^{2}}, \hspace{0.2cm}
F =\left(\frac{f+1}{2}\right)|F_0|
\label{Constantes}
\end{equation}
with,
\begin{equation}
\frac{1}{\tau_s}=\left(\frac{f-1}{2}\right)\frac{|F_0|}{T}
\end{equation}

The model given by Eq. (\ref{Ginzburg}) take the same form of the celebrated time-dependent Ginzburg-Landau equation \cite{Binney, Komin}, now added with stochastic noise \cite{Machado}. This coincides with the idea that most classical models for phase transitions are inherently nonlinear\cite{Yeomans} and at the same time, satisfies one of the conditions of SOC: non-linear interaction, normally in the form of thresholds \cite{Watkins}. In Eq. (\ref{Ginzburg}), the threshold transition parameter $T$ and the ratio $f$ control the time parameter $\tau_s$. This is a new characteristic time that competes with the damping time $\tau_0$. \par
Also, we can use the alternative SH precipitation model given by Eq. (\ref{eq:precipimodel2}).  Up to terms of order $q^{3}$, we  obtain a general model that contains the Ginzburg-Landau as a particular case,  
\begin{equation}
\begin{split}
\frac {\partial q}{\partial t} & = b\nabla ^{2}q +\frac{q}{\tau_s}+Gq^{2}-Kq^{3} \\ & +  D\dot{W} + F
\label{Ginzburg2}
\end{split}
\end{equation}
where $G$ defined as,
\begin{equation}
G = \frac{f-1}{2T\tau_0}, \hspace{0.2cm} 
\end{equation}
The main difference between Eq. (\ref{Ginzburg}) and (\ref{Ginzburg2}) is the quadratic term which vanishes in the Betts–Miller-like rain rate model, resulting in the Ginzburg-Landau equation. As is well known, the quadratic term  in the Ginzburg-Landau equation does not appear due to symmetry considerations. Here we will only study the Ginzburg-Landau equation, as the resulting pattern obtained from the second model were very different from real fields.\par

Fig. \ref{Phases} i)-l) shows the outcomes of the first model found solving numerically Eq. \ref{Ginzburg}. Further details of the simulations are explained in the Supplemental Material,  including several limiting cases studied to validate the software. It is worthwhile mentioning that the spectra in Fig. \ref{Phases} i)-l) were obtained from temporal averages once the system was relaxed to an stationary state. More structure is observed in the non-linear model when compared with the pure linear one. This is especially visible for intermediate regimes where the POCs are well defined. \par
As was done previously with the linear model, in the following section we further compare the outcomes of our non-linear model with the original clouds formations using Fourier spectrum and the closed-cell convection as reference. 
\subsection{Phase Transitions Diagrams}

\begin{figure}[t]
	\centering
	\includegraphics[width=1.0\linewidth]{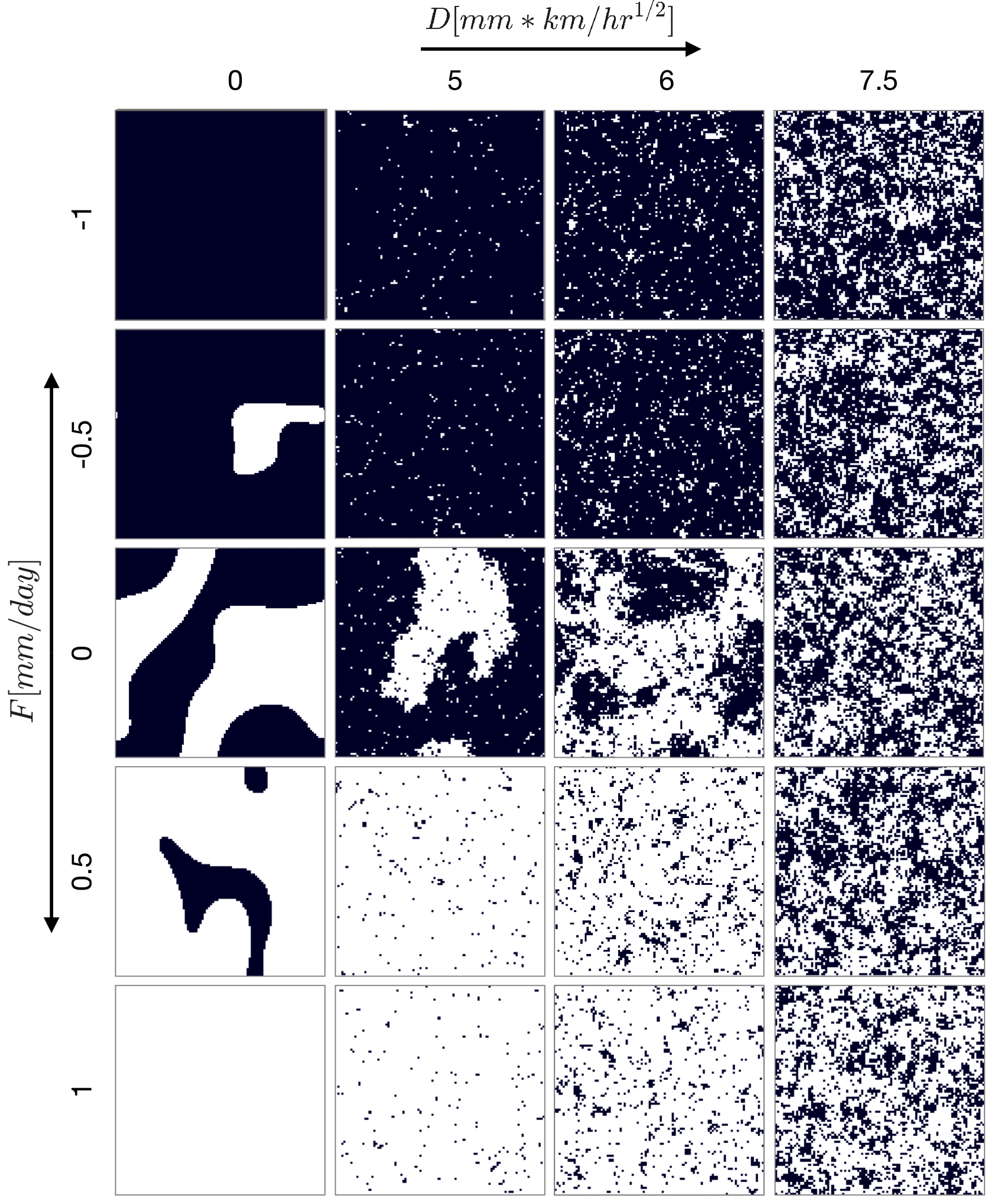}
	\caption{Representative patterns obtained as a function of the control parameters $D$ and $F$ for the stochastic Ginzburg-Landau equation.  For all the plots, we set $E= 1$ $hr^{-1}$ and  $K=  1$ $mm^2   \text{*}hr^{-1}$. Notice that $D$ and $F$ have values in the same range of  found by the original HS model from observational data \cite{Hottovy}.}
	\label{Fig:Phases}
\end{figure}

\begin{figure}[t]    \centering
	\includegraphics[width=1.0\linewidth]{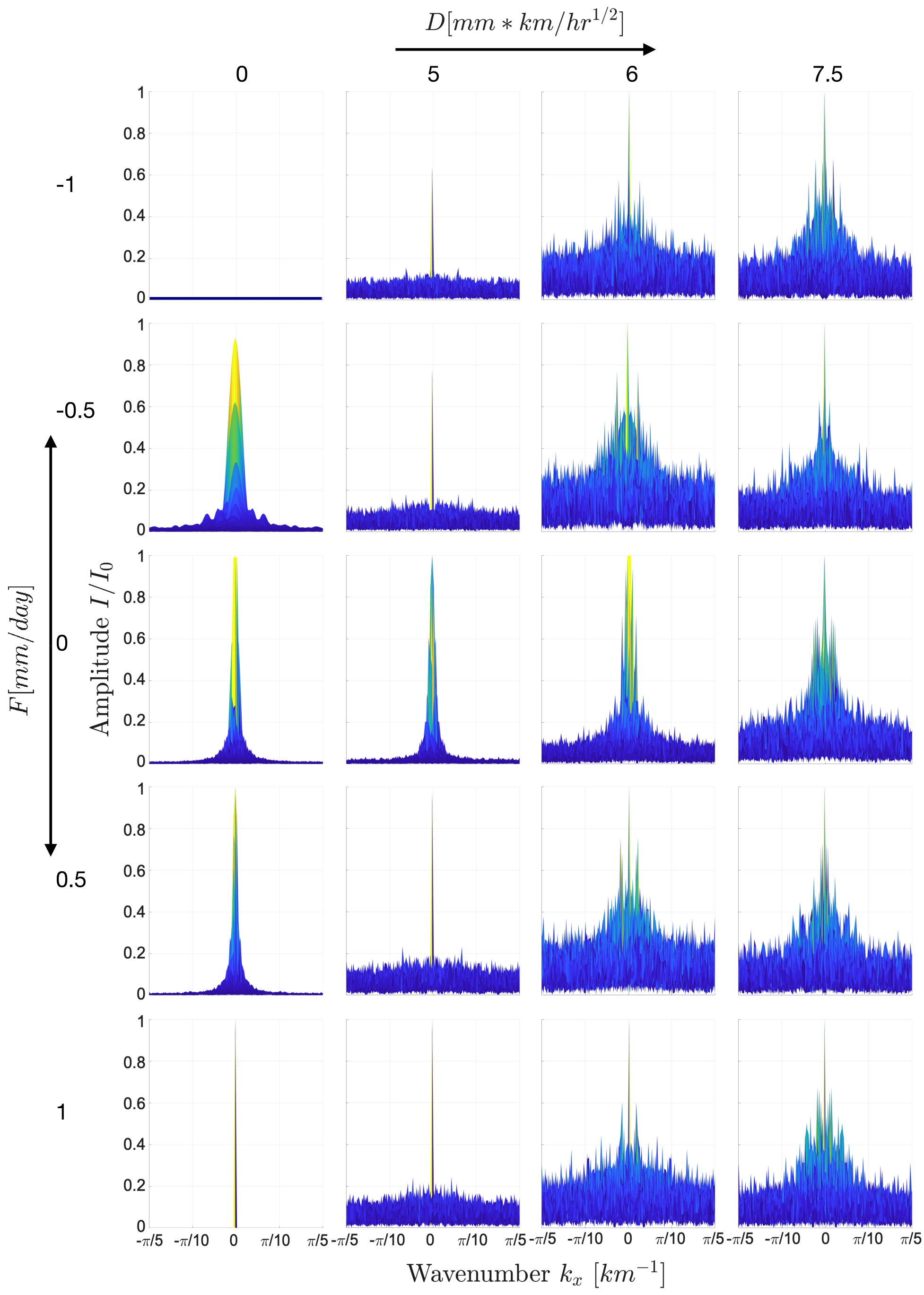}
	\caption{Fourier spectrum phase diagram  for the stochastic Ginzburg-Landau equation obtained as a function of the control parameters $D$ and $F$, using the patterns indicated in  Fig. \ref{Fig:Phases}. }
	\label{Fig:PatternsPhases}
\end{figure}

\begin{figure}[t]    \centering
	\includegraphics[width=1.0\linewidth]{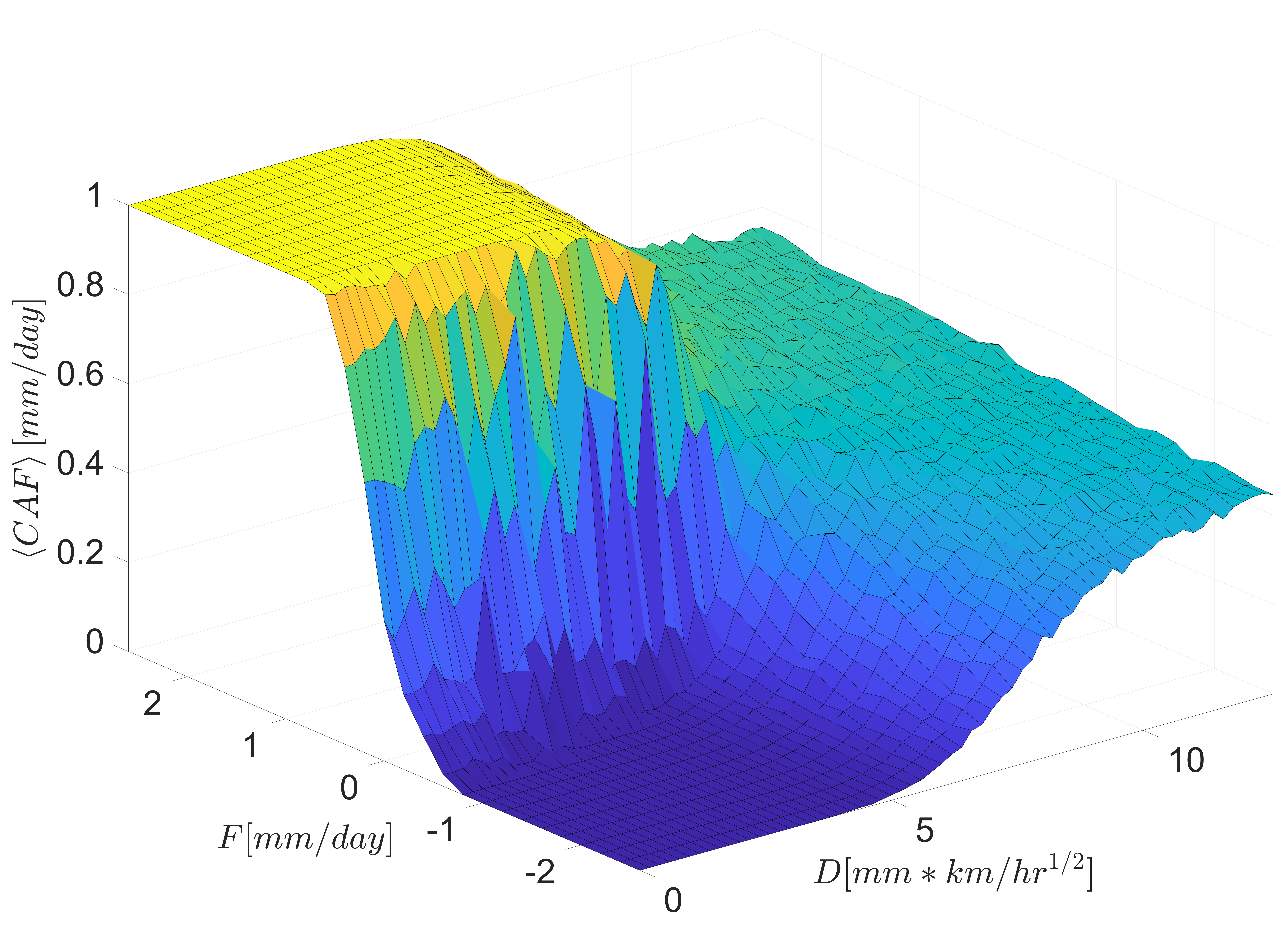}
	\caption{Phase diagram of shallow cloud regimes for the Ginzburg-Landau non-linear stochastic model given by Eq. (\ref{Ginzburg}). The plot shows the mean cloud area fraction ($\langle CAF \rangle$ ) as a function of variability, $D$, and the net source/sink parameter $F$. The transition from open to close cells is clearly seen as a transition from high to low values of the $\langle CAF \rangle $.  However, this picture changes by increasing $E$ and $K$,  resulting in two limiting cases (see Supplemental Material \cite{Supplemental}).}
	\label{Average}
\end{figure}

 The model outputs in Fig. \ref{Phases}, panels e)-h) present the four phases of cloud organization shown in observational data from panels a)-d), respectively. It is possible to see the transition from closed-cells to pockets of open cells (POCs). These four cloud regimes correspond to four distinct parameter regimes of Eq. (\ref{Ginzburg}) where $F$ and $D$ are the tuning parameters which determine the phase transition.\par
Fig. \ref{Fig:Phases} presents the phase diagram for different patterns, obtained from the stochastic Ginzburg-Landau equation, in
cases where they are qualitatively different as a function of the control parameters $D$ and $F$.
Fig. \ref{Fig:PatternsPhases} presents the Fourier spectrum of the corresponding patterns seen in Fig. \ref{Fig:Phases}. The control
parameter values are similar to those found in the HS model, obtained through a careful tuning of the model with real data \cite{Stechmann}. The
only difference here is the constants $E$ and $K$, which adjust the Fourier amplitude and position of the extra peaks. However, as explained in the supplementary material, these constants do not change for the different patterns, instead  were fixed at $E= 1$ $hr^{-1}$ and  $K=1$ $mm^2 \text{*} hr^{-1}$.\par
 Notice how in Fig. \ref{Fig:PatternsPhases}, for most of the patterns we are not able to see peaks other than the central one in the Fourier spectrum. These correspond to fields of the type shown in  \ref{Phases}, panels c)-d) which do not present much structure. Its Fourier spectrum is a bell-shaped curve centered at $k=0$, expected for such limiting cases. Other resulting patterns may have structure as in closed cells fields.\par
For example, Fig. \ref{Fig:FourierAndFields}k)-l) reveals the presence of a dominant frequency as observed in some real patterns Fig. \ref{Fig:FourierAndFields}b)-c). This kind of spectrum is radial symmetric, implying that the corresponding structure is glass-like, as it has a certain short range order but it is not preserved at long scales.\par

The peaks at $k\neq 0$ seen in Fig. \ref{Fig:FourierAndFields} h) are problematic to explain as the typical power spectrum for the stochastic Ginzburg-Landau or Cahn-Allen equation is a bell-shaped curve centered at $k=0$. Such peaks are usually only observed  under non-periodic boundary conditions or during transients. In fact, as shown in Figs. \ref{Fig:Phases} and \ref{Fig:PatternsPhases}, we reproduce a bell shaped curve in the region  where the noise can be taken as a small fluctuation in the Ginzburg-Landau equation, $i.e.$, for small $D$. As seen in the last column of Figs. \ref{Fig:Phases} and \ref{Fig:PatternsPhases}, in the limit where the noise starts to dominate, the fine structure of the potential washes away. The Fig. \ref{Fig:FourierAndFields} h) pattern lies in a special parameter region where noise and the non-linear functional power are of the same order. In noise sustained patterns as in adaptive control algorithms, this region turns out to be the most interesting as it contains a lot more "structural" information \cite{Leocher_2003}. As all benchmarks were reproduced in the limiting cases, including changes in the boundary conditions (see Supplementary Material), this means that either the state is stable or we have the following possibilities. One is that the system can be trapped in a deep metastable state. The other is a kind of numerical artifact.   It is well established numerically and mathematically that if the mesh size in the Ginzburg-Landau or Cahn-Allen equation simulation is shrunk, the numerical solutions would converge to a zero-distribution with no pattern formation in the continuum limit \cite{Sancho_1999,Ryser_2012}. In fact, the two-dimensional white noise-driven Allen–Cahn equation does not lead to the recovery of a physically meaningful limit \cite{Ryser_2012}. A way to interpret the simulations of such equation is to view them as numerical approximations of equations driven by a noise field having a finite correlation length\cite{Sancho_1999}. Here we used the mesh proposed by HS which has carefully tuned to reproduce meaningful physical results\cite{Hottovy}. However, we verified that the mesh only has a small effect in the peak position, as the mesh is associated with much higher values of $k$ and not at the center of the spectrum. \par
In fact, numerically such patterns appear for $E+F>0$ and its reason is easy to understand. The most simple analysis is obtained by linearization of the  average field $q=\langle q \rangle$ in Eq. (\ref{Ginzburg}), 
\begin{equation}
    \frac{\partial \langle q \rangle}{\partial t}=b\nabla^{2} \langle q \rangle+E\langle q \rangle+F
\end{equation}
Considering a field, $\langle q \rangle=\delta q \exp(i\boldsymbol{k}\cdot \boldsymbol{r}+\lambda t)$ results in the condition,
\begin{equation}
    \lambda=-bk^{2}+(E+F)
\end{equation}
The average field is stable whenever the real part of $\lambda$ is such that $Re(\lambda)=E+F<0$. 

Therefore, we conclude that either we are looking at a deep metstable state or there is a stable state with more structure. In the following subsection we further explore the pattern phase diagram of the system. \par
\begin{figure}[t]
	\centering
	\includegraphics[width=1.0\linewidth]{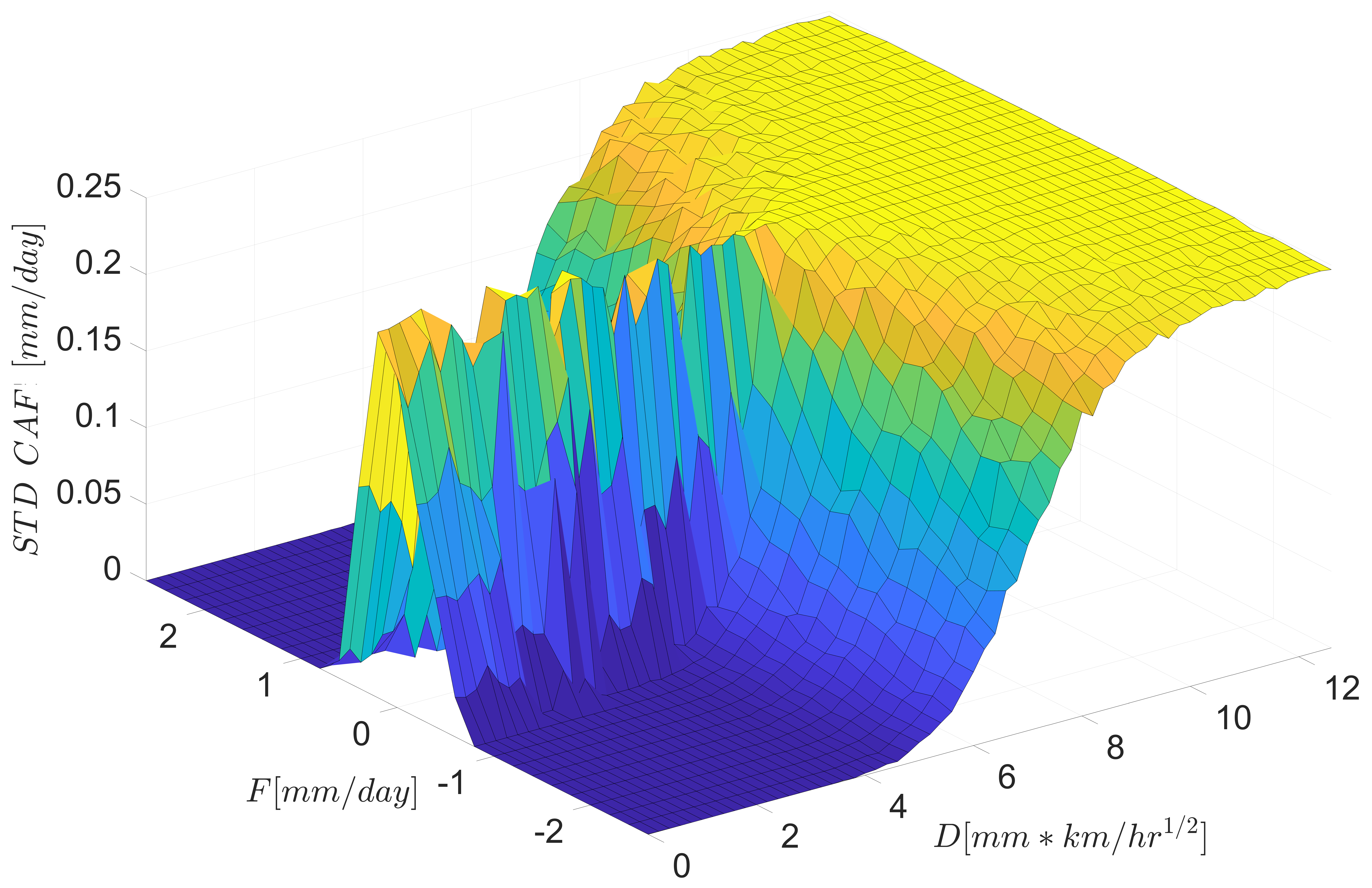}
	\caption{Plot of the cloud area fraction standard
		deviation  ($STDCAF$) as a function of the variability, $D$, and net source/sink, $F$, for the Ginzburg-Landau stochastic model given by Eq. (\ref{Ginzburg}). The open and closed cellular regimes are associated with low values of the $STDCAF$. The POCs and shallow phases are associated with high values of the $STDCAF$.}
	\label{Standar}
\end{figure}

To further understand the changes between one and another phase, we use a phase diagram of cloud regimes using statistics moments as shown in Figures \ref{Average} and \ref{Standar}. In the first diagram, the mean cloud area fraction ($\langle CAF \rangle$) is calculated as a function of $D$ and $F$, $i.e.$, $\langle{\sigma} \rangle = \langle {\sigma}(F,D) \rangle=\sum_{i,j}\sigma_{i,j}$ in the stationary state and by fixing $\tau_0$ and $b$. Moreover, the plot in Fig. \ref{Standar} provides  the standard deviation, which is a measure of the statistical sensitivity.\par
In Fig. \ref{Average} is notorious the phase diagram regions belonging to each regime: the closed-cell regime corresponds to $F >0$ and the open-cell regime corresponds to $F<0$, as indicated by the mean CAF, since while the average value cloud area of open cells is $1$, the mean of the closed ones is $0$. On the other hand, the POCs could be seen in the middle of both regimes as their transition in the region around $F=0$ with intermediate values of the mean $CAF$ between $0$ and $1$. All these cellular regimens are associated with intermediate values of $D$. The shallow cumulus regime (Fig.\ref{Phases}d) appears for $D> 8$ $mm*hr^{-1/2}$ at all $F$ values \cite{Supplemental}.\par

It is intuitive to understand why for small $D$, the CAF attains its mean unordered value: in this case, the value $\langle CAF \rangle=1$ should be reached for positive $F$, and $\langle CAF \rangle=0$ for
negative $F$. However, higher values of $E$ and $K$ affect this picture (see below).

Furthermore, to have a measure of the climate response or climate uncertainty, in Fig.\ref{Standar} we present the standard  deviation of the cloud area fraction  (STDCAF). The open and closed cellular regimes are associated with low values of the STDCAF. The POCs and shallow phases are associated with high values of the STDCAF, indicating how small changes in $F$ or $D$ lead to very large changes in $\langle CAF \rangle$. It also shows how the STDCAF increases drastically out of the regions where it presents the closed or open cellular patterns.

Finally, it's important to mention the effect of the $E$ and $K$ parameters on the phase diagram showed in Fig.\ref{Average}. After a systematic tuning, we observed a phase transition sensibility with respect to these parameters, $i.e.$, the change of the $E$ and $K$  values result in different phase spaces in which, even when it's possible to recover the four regimes of interest, the $F$ and $D$ couples able to form each phase suffer variations; in the Supplementary Material is discussed one example. On the other hand, fixing $F$ and $D$ at the values used for the cellular regimes, we conclude that even when these phases could be formed, the dominant amplitudes in their Fourier spectra changes for the effect of the $E$ and $K$  parameters. To conclude the physical interpretation of both parameters and their possible relevance in the clouds regimes formation it's necessary a further study.

\section{Stochastic Swift-Hohenberg model}
\label{sec:Swift}

\begin{figure}[t]
	\centering
	\includegraphics[width=1.0\linewidth]{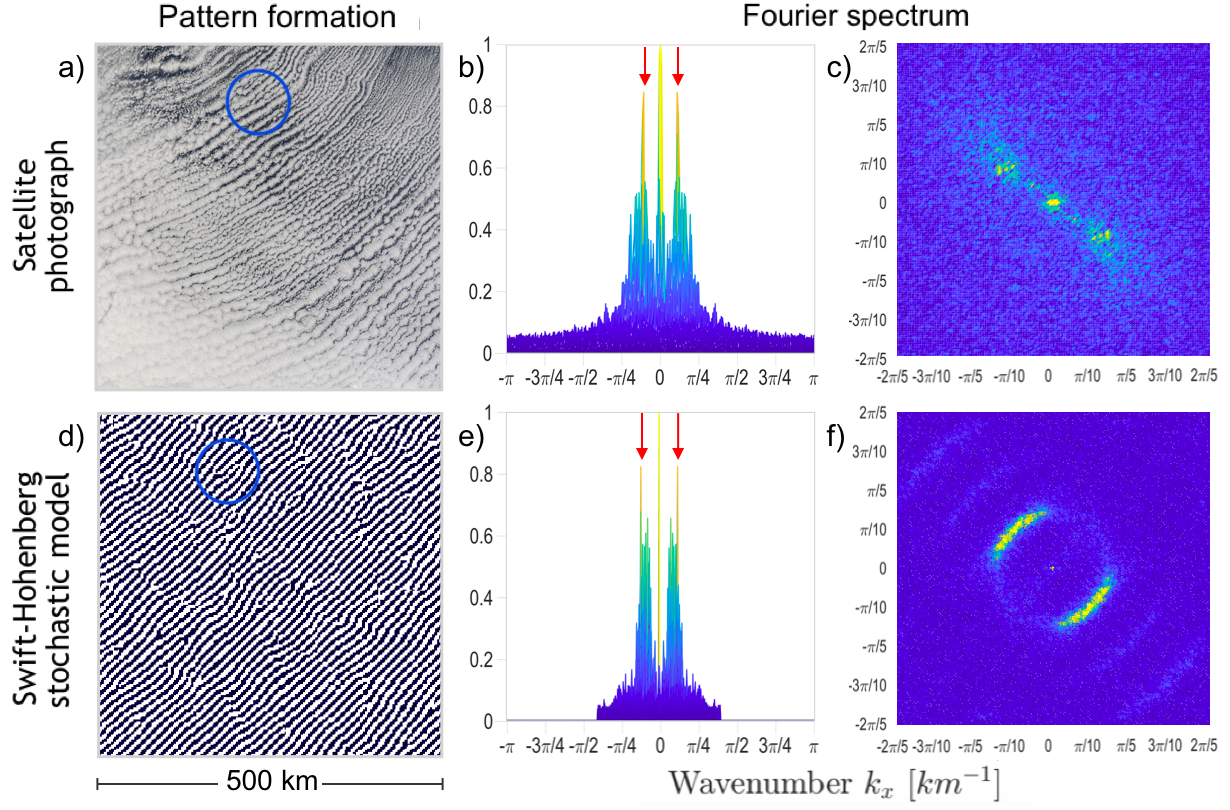}
	\caption{Fourier transform of the horizontal convective rolls. Panels in the left column show the horizontal convection pattern taken from a) satellite photograph and d) the Swift-Hohenberg model given by Eq. (\ref{eq:SHohenberg}).  In the central and right column are presented the Fourier spectra in the $I/I_0-k_y$ plane and in the orthogonal plane, respectively. We can identity in panels  b)-c) and e)-f) a dominant frequency with axial symmetry indicated by red arrows. Notice that in panels a) and d), the blue circles indicate bifurcations observed in the real and simulated patterns. See Supplemental Material for the parameter values \cite{Supplemental}.  The data of the real fields was taken from the Moderate Resolution Imaging Spectroradiometer (MODIS) data, and from the Geostationary Satellite Server (GOES) data from NOAA.}
	\label{Hohenberg}
\end{figure}

In spite that the stochastic non-linear models already show certain organization, Figs. \ref{Phases} a) and \ref{Hohenberg} a) reveal that some real cloud fields still can be much more organized and in fact are in a different physical limit. They reveal hexagonal cells mimicking patterns arising from Rayleigh-Bénard convection. Indeed they are considered as a form of Rayleigh-Bénard convection in moist atmospheric air \cite{Krishnamurti_1975,Feingold}. For such special clouds fields, we need to depart from some assumptions of the original HS model as after an exhaustive exploration of the parameters phase diagrams, there is no way to generate such highly ordered patterns. The dominant turbulent diffusion term prevents them to form. Returning to  the budget equation (\ref{eq:budget}), we see two possibilities. Either the source term or the small-scale flux convergence terms induce the selection of certain wavelengths. As clouds move, the pattern can persist in time, thus the source term is improbable to produce such behavior and we can keep our heuristically derived terms.   The next natural step is to consider changes in the small-scale flux convergence term, $i.e.$, in the operator $L(\Delta)$. The idea behind such change is the following. Suppose a pattern in which a wave-mode $\boldsymbol{k}_c$ is selected in an otherwise isotropic system. Let $\tilde{q}=\tilde{q}(\boldsymbol{k},t)$ be the Fourier transform of $q(\boldsymbol{r},t)$ in the space domain. The leading order dynamics must be of the following form,
\begin{equation}
\frac {\partial \tilde{q}}{\partial t}=(\alpha|\boldsymbol{k}|^{2}-\beta|\boldsymbol{k}|^{4}+...)\tilde{q}
\end{equation}
where  $\alpha>0$ and $\beta>0$, as we require small-wavelength modes to decay, i.e., $\tilde{q}(\boldsymbol{k},t)\rightarrow 0$ for $\boldsymbol{k}\rightarrow \infty$. In terms of the constants, the selected wave-mode is given by $k_c=\sqrt{\alpha/2\beta}$. Transforming to real space, we are lead to the following general equation,
\begin{equation}\label{eq:OriginalSH}
    \frac {\partial q}{\partial t}=-\alpha \nabla ^{2} q -\beta (\nabla ^{2})^{2}q
\end{equation}
We can take $\beta=1$ as scale. Therefore $\alpha=2k^{2}_c$ and we complete squares in Eq. (\ref{eq:OriginalSH}),
\begin{equation}
     \frac {\partial q}{\partial t}=(k_c^{4}-(k_{c}^{2} + \nabla^{2})^{2}) q
\end{equation}
This procedure to find the operator works for many types of pattern forming systems \cite{Croos3,Hecke}, but was first formally deduced from the Navier-Stokes equations in the Boussinesq approximation to study  the effects of
thermal fluctuations on a fluid near the Rayleigh-Benard instability \cite{Swift_1977}.  
By considering the expansion of $N\left[q\right]$ in Eq. (\ref{Ginzburg2}) and  collecting the linear terms in $q$ using a constant $\epsilon=k_c^{4}+1/\tau_s$, we obtain the following stochastic  equation,
\begin{equation}
\begin{split}
\frac {\partial q}{\partial t}  &= \left[\epsilon - (k_{c}^{2} + \nabla^{2})^{2}\right] q + G q^{2} \\ 
&- Kq^{3} + F + D\dot{W} 
\label{eq:SHohenberg}
\end{split}
\end{equation}

which is the stochastic Swift-Hohenberg equation. The  solutions of Eq. (\ref{eq:SHohenberg}) are still in the process of being investigated \cite{Gao_2017}  although  studies of the Swift-Hohenberg equation in the presence of noise started in the last decades \cite{PhysRevLett.71.1542}.  This is the general form, and probably the most simple model in the development of the Ginzburg-Landau theory of amplitude equations \cite{Doelman}. In fact, the Ginzburg-Landau model could be recovered by rescaling the long spatial and time scales \cite{Klepel,Hecke, Saarloos}.\par 
Eq. (\ref{eq:SHohenberg}) can be solved numerically through implicit finite differences and a successive over-relaxation (SOR) method as proposed by S. Sánchez Pérez-Moreno et al. \cite{Chavarria}. In Fig. \ref{Fig:FourierAndFields} j) and Fig. \ref{Hohenberg} d) we show the formation of two particular patterns that arise in the Rayleigh-Bénard convection, hexagons and rolls. Further details of the simulations are explained in the Supplemental Material \cite{Supplemental}. Both patterns have been identified as ways of organization in Sc clouds and their formation depends on the parameter $G$ that controls the strength of the quadratic nonlinearity. In Fig.\ref{Fig:FourierAndFields} panels a), j) and in Fig.\ref{Hohenberg} panels a),  d) we compare satellite photographs with simulations of hexagons and rolls, respectively; we can see clear similarities with the satellite patterns. To confirm the similarities, the Fourier spectrums of the real and simulated cloud formations were performed. \par
In Fig.\ref{Fig:FourierAndFields} panels b)-c) and k)-l), the hexagonal pattern spectrum reveals the presence of a dominant frequency for a cut along a certain direction.  In Fig.\ref{Fig:FourierAndFields} we can identify a principal frequency and other harmonics of lower amplitude. This coincides with the spectrum of a cellular pattern with defects and not highly ordered as a result of the forcing added in Eq. (\ref{eq:SHohenberg}), which generates different sizes of cells without a particular tessellation. On the other hand, in Fig.\ref{Hohenberg} panels b)-c) and e)-f)  we show the presence of a dominant frequency with axial symmetry that corresponds to a pattern formed by parallel rolls in real space. In both kinds of convection, the simulations recover the structures formed in real clouds fields.

\section{Conclusions}\label{sec:conclusions}
Following the work of Hottovy and Stechmann, we proposed a non-linear differential equation for an order parameter field given by the column water vapor $q(\mathbf {r} ,t)$ to describe the transitions of various pattern formations in mesoscale shallow clouds systems. One of the main modification introduced to the original linear model is the possibility of a feedback due to sources. In particular, we used two precipitation rate models, one leading to a time-dependent stochastic Ginzburg-Landau equation while the other adds a quadratic term to this equation.  The first model produces realistic cloud fields and even glass-like patterns, $i.e.$, with certain short range order which is not preserved at long scales.\par
However, this model is not able to reproduce the highly ordered fields present in Rayleigh-B\'enard convection in moist atmospheric  air of roll and hexagonal waves.  Therefore, following the theory of order parameter we introduced a change in the small-scale flux convergence term, resulting in a stochastic  Swift-Hohenberg equation, proposed here as a simple model for such clouds fields. The numerical simulations confirmed the presence of closed-cellular and horizontal convection phases.\par
The success of both models can be appreciated by observing the real patterns in Fig. \ref{Phases}. Therein, we identified that the three patterns corresponding to MCC are not in a perfectly hexagonal arrangement (highly ordered) nor are they arranged in complete randomness (highly disordered). The distributions of cumulus, both in closed and open-cells, appear in some arrangement between these two extremes.  \\
Both proposed non-linear models  are closer from this dominant structure that the linear one, while the Swift-Hohenberg equation allows the formation of patterns with a clear organization for two characteristic convective regimes. 
 Finally, we presented the phase diagram for the cloud patterns, using as basic parameters those found by HS by fitting the data, and the extra non-linear parameters found here by comparing with the space Fourier transform of the patterns.


\begin{acknowledgments}
	In Figs.\ref{Phases}, \ref{Fig:FourierAndFields} and \ref{Hohenberg}, the satellite images were taken from the Moderate Resolution Imaging Spectroradiometer (MODIS) data, available from NASA at https://earthobservatory.nasa.gov, and from the Geostationary Satellite Server (GOES) data from NOAA at https://www.nesdis.noaa.gov. \par
	We thank UNAM DGAPA-PROJECT IN102620. D. L.  Monroy thanks a  scholarship from DGAPA-UNAM. We also thank  Graciela B. de Raga (Instituto de Ciencias Atmosf\'ericas, UNAM, Mexico) and Michel Flores (Weizmann Institute, Israel) for sharing comments and clarifying certain points.  We also acknowledge helpful advice from Gerardo Ruiz-Chavarr\'ia (F. Ciencias, UNAM, Mexico)  on how to properly perform the simulations of the Swift-Hohenberg equation,  Aurora Hern\'andez-Machado (Universitat de Barcelona,Spain) and Denis Boyer (IFUNAM,Mexico) for providing critical advise and literature on the project.
	
\end{acknowledgments}

\providecommand{\noopsort}[1]{}\providecommand{\singleletter}[1]{#1}%
%

\pagebreak




\setcounter{section}{0}

\maketitle

\section*{SUPPLEMENTAL MATERIAL}
\subsection{Pattern Parameters}

\label{sec:Pattern_Parameters}
In this section, the domain and discretization, initial and boundary conditions, as well as the parameters values used in the numerical solutions of the models presented in the main text are explained in detail. For each cloud regime formed by the models, we also specify the tuning parameters that were used.

\subsubsection{The Stechmann and Hottovy linear Stochastic Model for mesoscale shallow patterns} 
In Fig. 1 panels a)-d) of the main text, the outcomes of the Eq. (3) were numerically solved using implicit finite differences with the same parameter values proposed by Hottovy and Stechmann \cite{Hottovy20,Stechmann20}. A two-dimensional discrete spatial grid in a domain of $L$ by $L$, where $L = 500$ $km$ divided in a $N$ by $N$ lattice with $N = 100$ and lattice spacing of $\Delta x = \Delta y = 5$ $km$; this was chosen to be roughly the smallest width of individual cells of tropical deep convection. The boundary and initial conditions were considered as periodic and random, respectively.  It was defined $q_{i,j}(t)$ as the integrated CWV and $W_{i,j}(t)$ as the independent white noises, denoted formally as the derivative of a Wiener process \cite{Hottovy20,Stechmann20}, in the $(i, j)th$ column of the atmosphere for $i, j = 1,..., N$. \\

The parameters $b$ and $\tau_0$ conserves the values $b=25$ $mm^{2}*hr^{-1}$ and $\tau_0 = 100$ $hr$ proposed in \cite{Hottovy,Stechmann}. In each phase of Fig. 1, the parameter values used were a) $D = 1.55$ $mm*hr^{-1/2}$, $F = 0.12$ $mm*day^{-1}$, b) $D = 1.94$ $mm*hr^{-1/2}$, $F = 0.048$ $mm*day^{-1}$ c) $D = 1.55$ $mm*hr^{-1/2}$, $F = -0.12$ $mm*day^{-1}$ and d) $D = 11.62$ $mm*hr^{-1/2}$, $F = -0.72$ $mm*day^{-1}$.\par

\subsubsection{Non-linear model: time-dependent Ginzburg-Landau stochastic equation}

In Fig. 1 panels i)-j), the outcomes of the Eq. (11) used the same domain and discretization as well as initial and boundary conditions of the linear model simulations.
The parameters $b$ and $\tau_0$ conserves the same value proposed by Hottovy and Stechmann \cite{Hottovy20,Stechmann20}, while different values of $F$ and $D$, in the same range used by them ($F_0\sim \pm1$ $mm*day^{-1}$ and $D\sim10$ $mm*hr^{-1/2}$), were explored to find the regimens observed in Fig. 1, panels i)-l). The dynamics of the non-linear terms in Eq. (11) was determined by the parameters $E$ and $K$ whose values, after an exploration of different orders of magnitude, were fixed in $E = 1$ $hr^{-1}$ and $K=1$ $mm^{2}*hr^{-1}$. The increase of both parameters is associated with a major percolation in the boundaries around open or closed clusters to the same $F$ and $D$ values. \\

In particular, the parameter values used in Fig. 1 for Eq. (12) were i) $D = 8.5$ $mm*hr^{-1/2}$, $F = 1$ $mm*day^{-1}$, j) $D = 9$ $mm*hr^{-1/2}$, $F = 0.2$ $mm*day^{-1}$ k) $D = 8.55$ $mm*hr^{-1/2}$, $F = -1$ $mm*day^{-1}$ and l) $D = 10.25$ $mm*hr^{-1/2}$, $F = -0.4$ $mm*day^{-1}$.\par

\subsubsection{Stochastic Swift-Hohenberg model}

In Fig. 2 g) and Fig. 5 c) we show the formation of two particular patterns that arise in the Rayleigh-Bénard convection, hexagons and rolls. Eq. (21) was solved numerically through implicit finite differences and a successive over-relaxation (SOR) method as proposed by S. Sánchez Pérez-Moreno et al. \cite{Chavarria20}.\par
For the simulations showed, the numerical method used a two-dimensional discrete spatial grid in a domain of $L$ by $L$, where $L = 500$ $km$ was divided in a $N$ by $N$ lattice with $N = 200$ and lattice spacing of $\Delta x = \Delta y = 2.5$ $km$. In this case, this discretization was chosen to approximate the cell diameter of the real ones. The boundary and initial conditions were considered again as periodic and random. In the SOR method, it was used as the iteration step $k = 15$ and as the relaxation factor $w=1.3$. \\

To form each pattern, the parameters were fixed as follows: in Fig.2 g)  $\epsilon = 0.1 $, $k_c = 1.3$ $m^{-1}$, $g=1$, $D = 0.15$ $mm*hr^{-1/2}$, $F = 0.1$ $mm*day^{-1}$. and in Fig.5 c) 
$\epsilon = 0.3 $, $k_c = 1.2$ $m^{-1}$, $g=0$, $D = 0.3$ $mm*hr^{-1/2}$, $F = 0.25$ $mm*day^{-1}$

\subsection{Fourier Transform Analysis}

To investigate the validity and accuracy of the main text results, in this section we present first, in subsections A, B and C, the Fourier Transform software testing and second, in section D, an examination of the numerical method used to solve the Eq. (11) varying the mesh grid and boundary conditions.

\subsubsection{Fourier Transform Benchmarks}

\begin{figure}[b]
	\centering
	\includegraphics[width=1.0\linewidth]{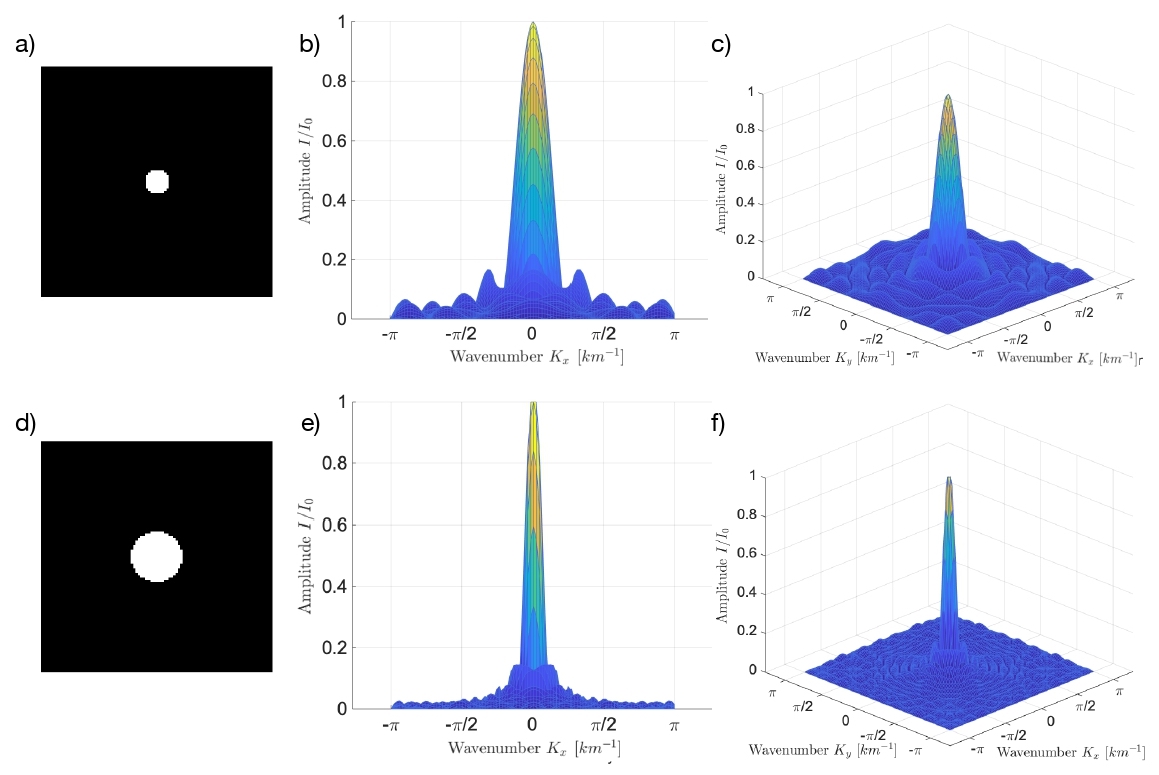}
	\caption{ Fourier transform code proof. a) Circle with radio $r =20$ pixels in a $200\times200$ square lattice, the corresponding 2D Fourier transform is showed in b) the $I/I_0-k_y$ plane and c) the orthogonal plane. d), b) and e) are equivalent to a), b) and c) to a circle with radio $r = 50$ pixels in a $200 \times 200$ square lattice.}
	\label{Fig:FourierTest}
\end{figure}

First we tested the Fourier spectrum software  using known examples to reproduce the expected results. Among the targets, the most simple one is two circular apertures with different diameters, as shown in \ref{Fig:FourierTest} panels a) and d). In the middle and right columns, the respective Fourier spectrum of each aperture is showed in b), e) the $I/I_0-k_y$ plane, and in c), f) the orthogonal plane. The analysis is in
perfect agreement with the expected analytical results. \par

\subsection{Time-averaged Fourier Transform}

\begin{figure}[t]
	\centering
	\includegraphics[width=1.0\linewidth]{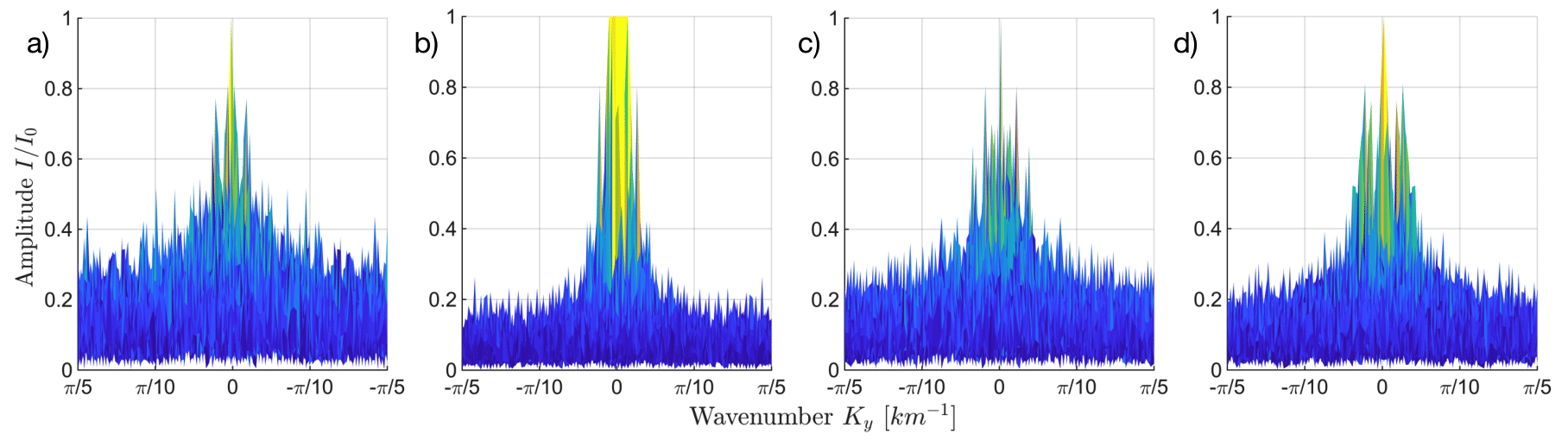}
	\caption{ The corresponding time-averaged Fourier transforms for the four distinctive phases of shallow cloud organization generated by the Ginzburg-Landau stochastic model (Eq. (11) with the same order and parameters used in Fig. \ref{Fig:Hott-Ginz}, panels e)-h). The Fourier transforms were averaged over 20 independent simulations in which the patterns present a stationary behavior (from $t=100$ to $t=300$ each 10 time-steps).}
	\label{Fig:Fourier-Average}
\end{figure}

We next investigate the persistence of the dominant wavenumbers that appears in the Ginzburg-Landau Fourier spectra. With this purpose, we computed the time-averaged Fourier spectra of the four distinctive cloud phases generated by the Ginzburg-Landau stochastic model (see Fig. \ref{Fig:Fourier-Average}) once the patterns reach a stationary state. \\

The corresponding 2D Fourier Transform of each phase was averaged over $20$ independent simulations in the total period $\lbrack 150, 350 \rbrack$ $hrs$ at time intervals of $t_n = 10*n + T_i$ for $n$ an integer. The beginning time, $T_i =  150$ $hrs$, corresponds to the common minimum time in which the four phases reach the stability according to the $\langle q (\mathbf {r} ,t) \rangle$  value. \\

We conclude that these characteristic wavenumbers are persistent in the stationary cellular phases. Maybe they could correspond to metastable states with a long decay time. We recognize that this is possibly equivalent to metastability presented in the two-dimensional Ising model under the effects of an external magnetic field.

\subsubsection{Comparison between Hottovy-Stechmann and Ginzburg-Landau Fourier spectra}

\begin{figure}[h]
	\centering
	\includegraphics[width=1.0\linewidth]{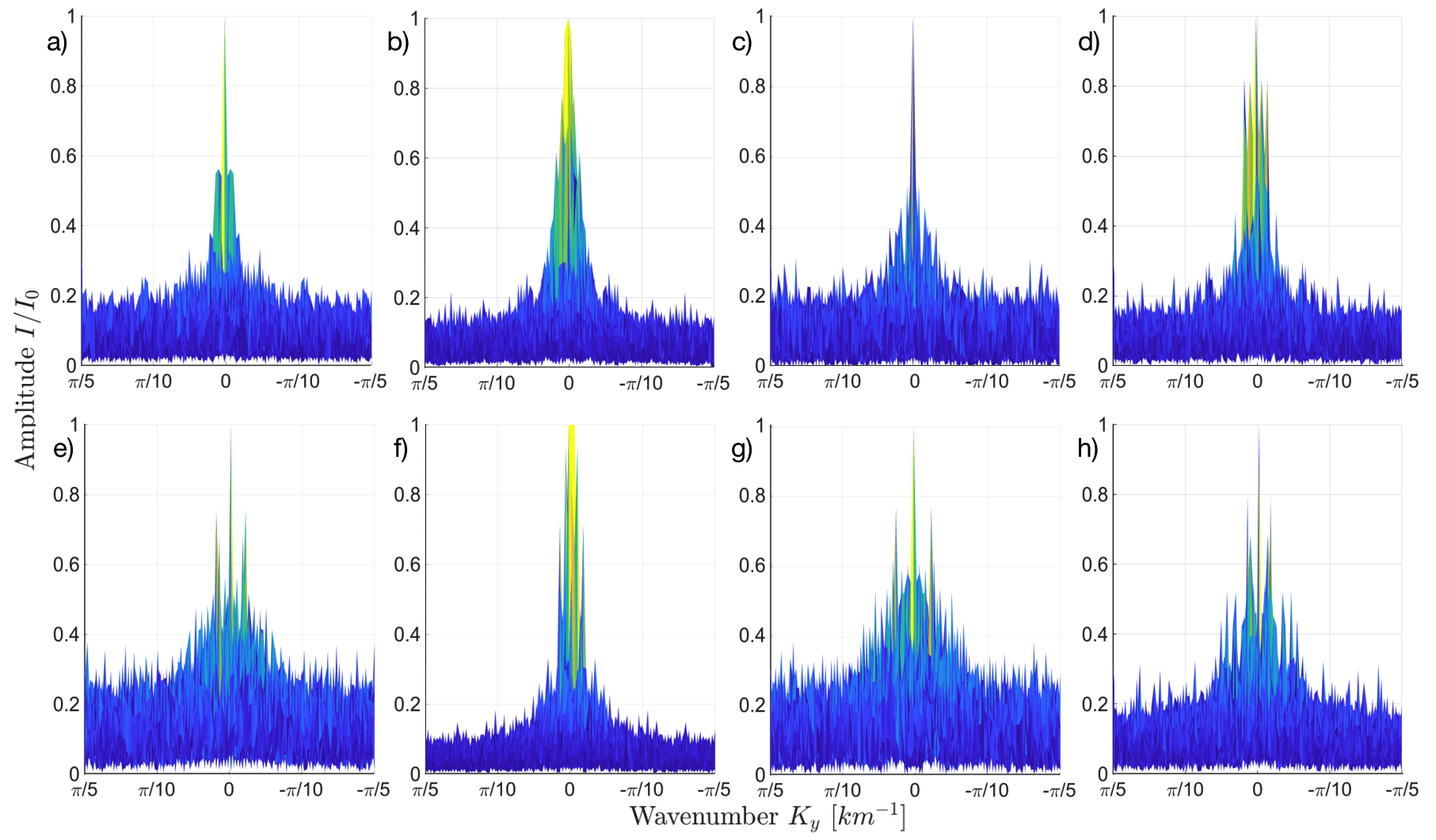}
	\caption{The corresponding Fourier transforms for the four distinctive phases of shallow cloud organization: closed-cell stratocumulus,  pockets of open-cell stratocumulus,  open-cell stratocumulus, and shallow cumulus generated by the HS model (Eq. (3)) with the parameters proposed in Ref \cite{Hottovy20} in panels a) to d) and by the Ginzburg-Landay model (Eq. (11)) in panels e) to h). See section \ref{sec:Pattern_Parameters} for the parameter values.}
	\label{Fig:Hott-Ginz}
\end{figure}

Once we had evidence of the Fourier spectra validity used in the patterns analysis, we investigate the role of non-linear terms of the Ginzburg-Landau model in the emergence of patterns for certain couples of F and D parameter values. Fig. \ref{Fig:Hott-Ginz} shows the Fourier transform corresponding to the four cloud phases of interest generated by the Hottovy-Stechmann model, in panels a-d) and by the Ginzburg-Landau model, in panels e-h).\\

The Fourier transforms in the top row show no dominant wave numbers over the rest, which is consistent with the lack of organization in the Hottovy-Stechmann patterns. However, the bottom row presents, as the Fig. \ref{Fig:Fourier-Average} does, characteristic wave numbers that give a first clue of a more homogeneous distribution and so, more organization in the patterns formed by the Ginzburg-Landau model. Also, the fact that these dominant wave numbers appear only in the cellular phases allows to complement the phase diagrams in the main text to understand the effect of the tuning parameters, $F$ and $D$, in the formation and transition of cloud phases.

\subsubsection{Characterizing the effect of boundary conditions and mesh grid on Ginzburg-Landau Fourier spectra}

Most of the numerical studies which have been found  disordered spatio-temporal regimes formed by non-linear partial differential equations have been done  considering periodic boundary conditions, with the idea that in the limit of very large systems, the boundary conditions would not influence the system dynamics. However, for the description of real systems, it is necessary a systematic study of more complex boundary conditions to consider their possible effects in the formation of more realistic patterns \cite{Piro}. \\

For this reason, once we prove the validity of the Fourier Transform program as well as the numerical solution of Ginzburg-Landau model, in this section we will focus on the behavior of the stochastic Ginzburg-Landau equation on different mesh refinement and with different types of boundary conditions.\\

\begin{figure}[h]
	\centering
	\includegraphics[width=1.0\linewidth]{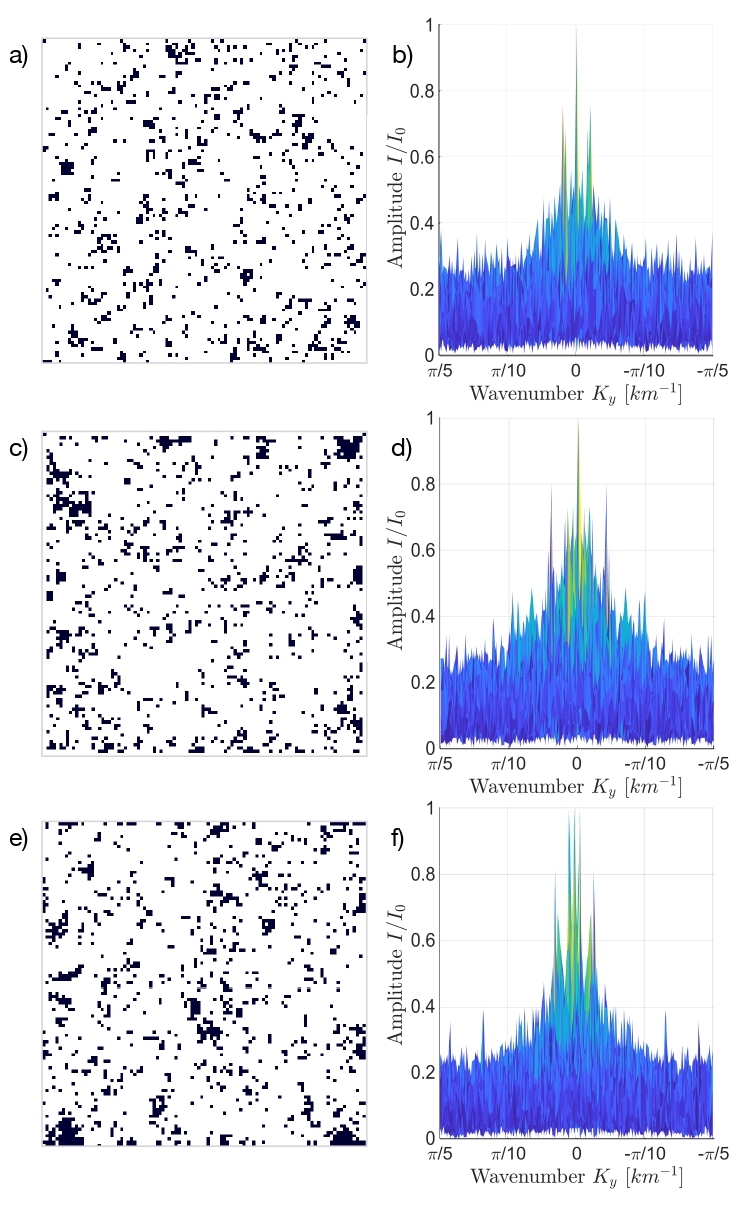}
	\caption{ Fourier transforms of the closed-cell phase. Panels in the left column show the closed cellular pattern taken from Ginzburg-Landau stochastic model (Eq. (11) using a) periodic boundary conditions, c) Dirichlet boundary conditions and e) Neumann boundary conditions. In the right column, we present the corresponding Fourier spectrum of each pattern. }
	\label{Fig:Fourier-Bound}
\end{figure}

First, through the comparison of periodic, Neumann and Dirichlet boundary conditions (see Fig. \ref{Fig:Fourier-Bound}) we summarize the behavior observed numerically on the closed-cellular regimen formed in a two-dimensional rectangular domain under the same parameters detailed in  section \ref{sec:Pattern_Parameters}.  \\

Applying null Dirichlet ($q=0$), and Neumann ($\frac{\partial q}{\partial n } = 0$) boundary conditions, in the left column of Fig. \ref{Fig:Fourier-Bound} we show the patterns formed under each kind of condition. In the right column we can see their respective Fourier spectra. For the three cases, the spectra reveal similarities between them. In the left panels, it is possible to appreciate such behavior qualitatively. However, for the Dirichlet and Neumann cases, near to the walls, we can see open regions in contradistinction to the periodic case. This behavior could be associated, in the Dirichlet case, with zero amplitude boundaries that facilitate the formation of defects and, in the Neumann case, for the absorption of the defects by the boundaries. In both conditions, the interference of the plane waves emitted by the walls is determinant in the pattern evolution \cite{Piro}.\\

\begin{figure}[h]
	\centering
	\includegraphics[width=1.0\linewidth]{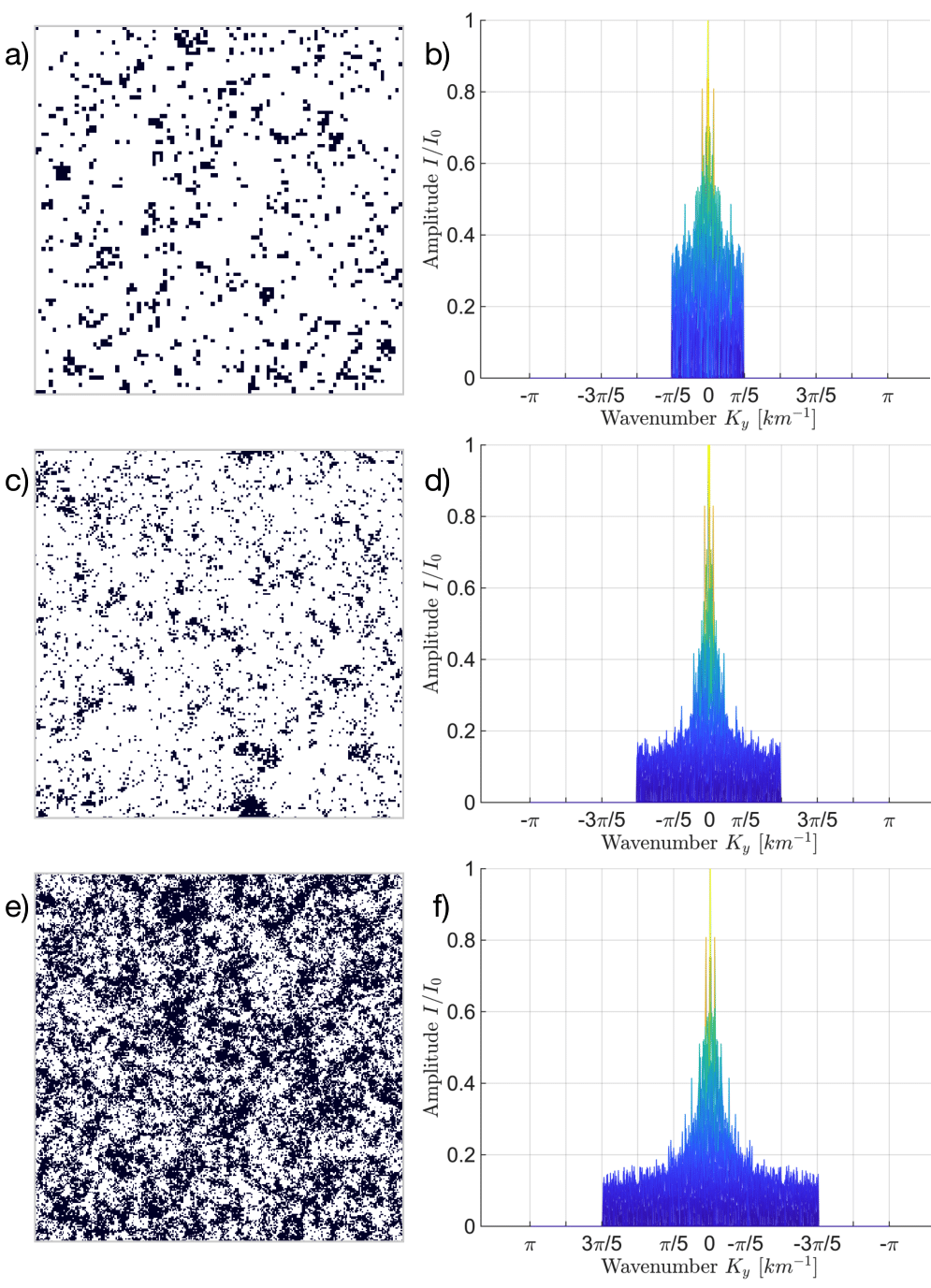}
	\caption{Fourier transform of the closed-cell phase. Panels in the left column show the closed cellular pattern taken from Ginzburg-Landau stochastic model (Eq. (11) solved in a square discrete domain of $L$ by $L$, with $L = 500$ $km$, divided in a $N \times N$ lattice with a) $N = 100$, b)$N = 200$ and $N = 300$.  In the right column, we present the corresponding Fourier spectrum of each pattern. The maximal spatial frequencies in panels b), d) and e) are determined by the resolution of the grid used in the simulation given in the units of $k_x$ (see main text).}
	\label{Fig:Fourier-Mesh}
\end{figure}

On the other hand, to investigate the effects of the mesh refinement on pattern formation, we simulate our system with the same initial and periodic boundary conditions specified in section \ref{sec:Pattern_Parameters} over a square domain with side $ L = 500$ km . In Fig. \ref{Fig:Fourier-Mesh} we present the results for different mesh refinements $ \Delta (x) = L / N $ where $N$ is the number of lateral divisions. In a), $ N = 100$, c) $ N = 200$ and e) $ N =  300 $ cells. By observation of the left column is clear that $ \Delta (x) $ affects the $ CAF $; particularly, in panel e), this is visible with the apparition of open regions and the decrease of the closed area percolation, compared with panels a) and c). Such effect has been reported previously by HS and that's why one need to  tune $ \Delta (x) $ with observational data.
In spite of this, our Fourier spectra results in the right column suggests a common behavior of the three patterns as similar dominant wave-numbers are visible.

\subsection{Ginzburg-Landau phase diagrams}

\begin{figure}[H]    \centering
	\includegraphics[width=.88\linewidth]{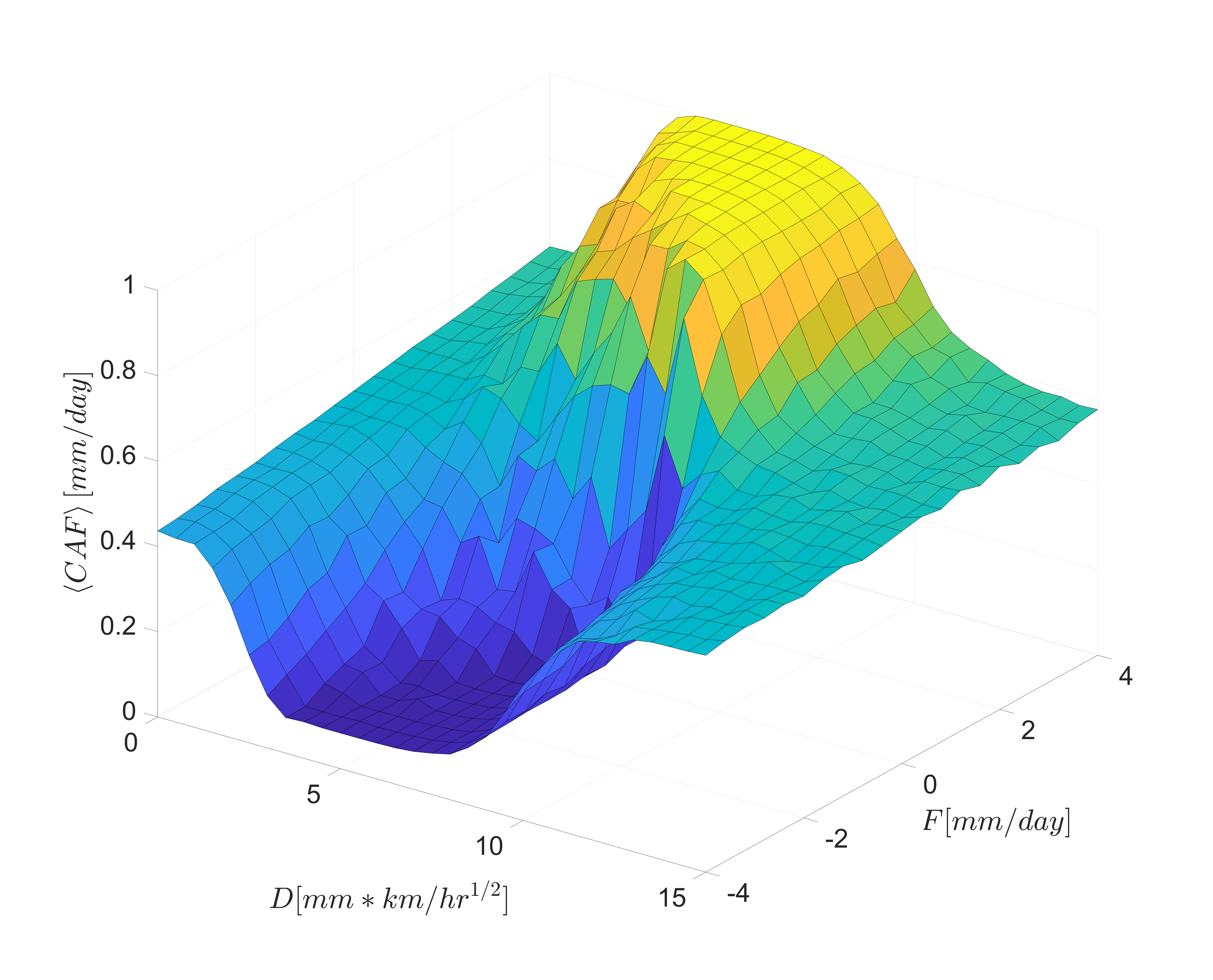}
	\caption{ Phase diagram of shallow cloud regimes for the Ginzburg-Landau stochastic model given by Eq. (11). The plot shows the mean cloud area fraction ($\langle CAF \rangle$) as a function of $D$ and $F$ fixing the parameters $E = 8.5$ $hr^{-1}$ and  $K= 6.5$ $mm^2*hr^{-1}$.}
	\label{Average2}
\end{figure}

\begin{figure}[H]
	\centering
	\includegraphics[width=.88\linewidth]{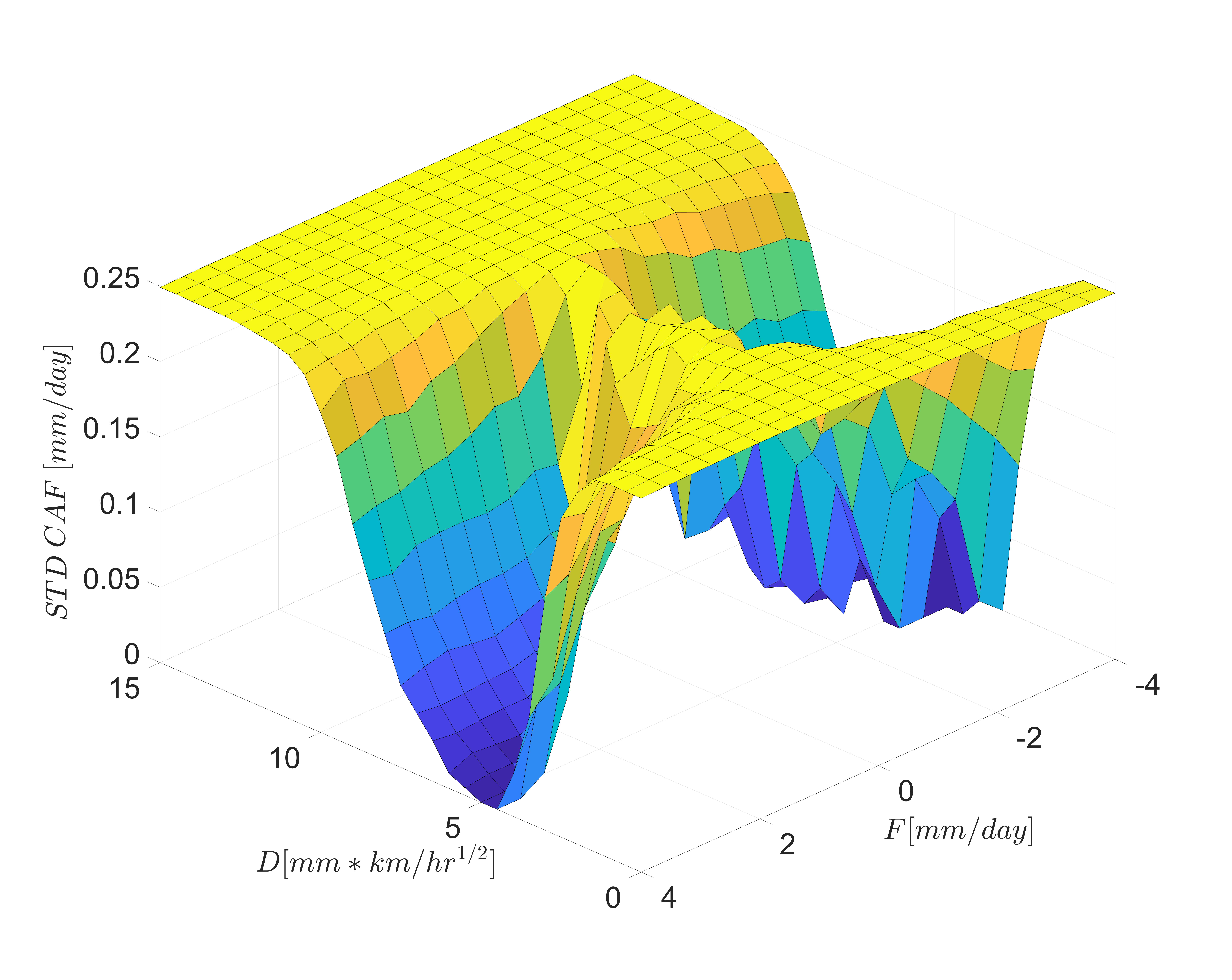}
	\caption{ Plot of the cloud area fraction standard
		deviation ($STDCAF$) as a function of the $D$ and $F$, for the Ginzburg-Landau stochastic model given by Eq. (11), fixing the parameters $E = 8.5$ $hr^{-1}$ and  $K= 6.5$ $mm^2*hr^{-1}$.}
	\label{Standar2}
\end{figure}

The study of the Ginzburg-Landau time-dependent equation requires to consider the effects of the linear and non-linear parameters in the phase formation and transition. Represented in the main text as $E$ and $K$, the polynomial terms in  Eq. (11)  were explored systematically by identifying two limits: 1) when $E$ and $K$ tend to 0 with results close to the Hottovy and Stechmann outputs and, 2) when $E$ and $K$ increase.  In the phase diagram, this produce the formation of symmetry with respect to an intermediate  $D$ value, as is shown in Figs. \ref{Average2} and \ref{Standar2}.\\

\color{black}


\begin{thebibliography}{57}%
	\makeatletter
	\providecommand \@ifxundefined [1]{%
		\@ifx{#1\undefined}
	}%
	\providecommand \@ifnum [1]{%
		\ifnum #1\expandafter \@firstoftwo
		\else \expandafter \@secondoftwo
		\fi
	}%
	\providecommand \@ifx [1]{%
		\ifx #1\expandafter \@firstoftwo
		\else \expandafter \@secondoftwo
		\fi
	}%
	\providecommand \natexlab [1]{#1}%
	\providecommand \enquote  [1]{``#1''}%
	\providecommand \bibnamefont  [1]{#1}%
	\providecommand \bibfnamefont [1]{#1}%
	\providecommand \citenamefont [1]{#1}%
	\providecommand \href@noop [0]{\@secondoftwo}%
	\providecommand \href [0]{\begingroup \@sanitize@url \@href}%
	\providecommand \@href[1]{\@@startlink{#1}\@@href}%
	\providecommand \@@href[1]{\endgroup#1\@@endlink}%
	\providecommand \@sanitize@url [0]{\catcode `\\12\catcode `\$12\catcode
		`\&12\catcode `\#12\catcode `\^12\catcode `\_12\catcode `\%12\relax}%
	\providecommand \@@startlink[1]{}%
	\providecommand \@@endlink[0]{}%
	\providecommand \url  [0]{\begingroup\@sanitize@url \@url }%
	\providecommand \@url [1]{\endgroup\@href {#1}{\urlprefix }}%
	\providecommand \urlprefix  [0]{URL }%
	\providecommand \Eprint [0]{\href }%
	\providecommand \doibase [0]{https://doi.org/}%
	\providecommand \selectlanguage [0]{\@gobble}%
	\providecommand \bibinfo  [0]{\@secondoftwo}%
	\providecommand \bibfield  [0]{\@secondoftwo}%
	\providecommand \translation [1]{[#1]}%
	\providecommand \BibitemOpen [0]{}%
	\providecommand \bibitemStop [0]{}%
	\providecommand \bibitemNoStop [0]{.\EOS\space}%
	\providecommand \EOS [0]{\spacefactor3000\relax}%
	\providecommand \BibitemShut  [1]{\csname bibitem#1\endcsname}%
	\let\auto@bib@innerbib\@empty
	\bibitem [{\citenamefont {Schneider}\ \emph {et~al.}(2017)\citenamefont
		{Schneider}, \citenamefont {Teixeira}, \citenamefont {Brient}, \citenamefont
		{Pressel}, \citenamefont {Schär},\ and\ \citenamefont
		{Siebesma}}]{Schneider}%
	\BibitemOpen
	\bibfield  {author} {\bibinfo {author} {\bibfnamefont {T.}~\bibnamefont
			{Schneider}}, \bibinfo {author} {\bibfnamefont {J.}~\bibnamefont {Teixeira}},
		\bibinfo {author} {\bibfnamefont {C.~S. B.~F.}\ \bibnamefont {Brient}},
		\bibinfo {author} {\bibfnamefont {K.~G.}\ \bibnamefont {Pressel}}, \bibinfo
		{author} {\bibfnamefont {C.}~\bibnamefont {Schär}},\ and\ \bibinfo {author}
		{\bibfnamefont {A.~P.}\ \bibnamefont {Siebesma}},\ }\bibfield  {title}
	{\bibinfo {title} {Climate goals and computing the future of clouds},\ }\href
	{https://doi.org/10.1038/nclimate3190} {\bibfield  {journal} {\bibinfo
			{journal} {Nature Climate Change}\ }\textbf {\bibinfo {volume} {7(1)}},\
		\bibinfo {pages} {3–5} (\bibinfo {year} {2017})}\BibitemShut {NoStop}%
	\bibitem [{\citenamefont {Nuijens}\ and\ \citenamefont
		{Siebesma}(2019)}]{Nuijens}%
	\BibitemOpen
	\bibfield  {author} {\bibinfo {author} {\bibfnamefont {L.}~\bibnamefont
			{Nuijens}}\ and\ \bibinfo {author} {\bibfnamefont {A.~P.}\ \bibnamefont
			{Siebesma}},\ }\bibfield  {title} {\bibinfo {title} {Boundary layer clouds
			and convection over subtropical oceans in our current and in a warmer
			climate},\ }\href {https://doi.org/10.1007/s40641-019-00126-x} {\bibfield
		{journal} {\bibinfo  {journal} {Current Climate Change Reports}\ }\textbf
		{\bibinfo {volume} {5(2)}},\ \bibinfo {pages} {80–94} (\bibinfo {year}
		{2019})}\BibitemShut {NoStop}%
	\bibitem [{\citenamefont {Deng}\ \emph {et~al.}(2002)\citenamefont {Deng},
		\citenamefont {Seaman},\ and\ \citenamefont {Kain}}]{Deng}%
	\BibitemOpen
	\bibfield  {author} {\bibinfo {author} {\bibfnamefont {A.}~\bibnamefont
			{Deng}}, \bibinfo {author} {\bibfnamefont {N.~L.}\ \bibnamefont {Seaman}},\
		and\ \bibinfo {author} {\bibfnamefont {J.~S.}\ \bibnamefont {Kain}},\
	}\bibfield  {title} {\bibinfo {title} {A shallow-convection parameterization
			for mesoscale models. part i: Submodel description and preliminary
			applications},\ }\href@noop {} {\bibfield  {journal} {\bibinfo  {journal}
			{Journal of the Atmospheric Sciences}\ }\textbf {\bibinfo {volume} {60}},\
		\bibinfo {pages} {34–56} (\bibinfo {year} {2002})}\BibitemShut {NoStop}%
	\bibitem [{\citenamefont {Stephan}\ \emph {et~al.}(2019)\citenamefont
		{Stephan}, \citenamefont {Hauke}, \citenamefont {Sandrine},\ and\
		\citenamefont {Bjorn}}]{Rasp}%
	\BibitemOpen
	\bibfield  {author} {\bibinfo {author} {\bibfnamefont {R.}~\bibnamefont
			{Stephan}}, \bibinfo {author} {\bibfnamefont {S.}~\bibnamefont {Hauke}},
		\bibinfo {author} {\bibfnamefont {B.}~\bibnamefont {Sandrine}},\ and\
		\bibinfo {author} {\bibfnamefont {S.}~\bibnamefont {Bjorn}},\ }\bibfield
	{title} {\bibinfo {title} {Combining crowd-sourcing and deep learning to
			understand meso-scale organization of shallow convection},\ }\href@noop {}
	{\bibfield  {journal} {\bibinfo  {journal} {Bull. Amer. Meteor. Soc.}\ ,\
			\bibinfo {pages} {1}} (\bibinfo {year} {2019})}\BibitemShut {NoStop}%
	\bibitem [{\citenamefont {Dagan}\ \emph {et~al.}(2018)\citenamefont {Dagan},
		\citenamefont {Koren}, \citenamefont {Altaratz1},\ and\ \citenamefont
		{Feingold}}]{Dagan}%
	\BibitemOpen
	\bibfield  {author} {\bibinfo {author} {\bibfnamefont {G.}~\bibnamefont
			{Dagan}}, \bibinfo {author} {\bibfnamefont {I.}~\bibnamefont {Koren}},
		\bibinfo {author} {\bibfnamefont {O.}~\bibnamefont {Altaratz1}},\ and\
		\bibinfo {author} {\bibfnamefont {G.}~\bibnamefont {Feingold}},\ }\bibfield
	{title} {\bibinfo {title} {Feedback mechanisms of shallow convective clouds
			in a warmer climate as demonstrated by changes in buoyancy},\ }\href@noop {}
	{\bibfield  {journal} {\bibinfo  {journal} {Environmental Research Letters}\
		}\textbf {\bibinfo {volume} {13(5)}} (\bibinfo {year} {2018})}\BibitemShut
	{NoStop}%
	\bibitem [{\citenamefont {Ceppi}\ \emph {et~al.}(2017)\citenamefont {Ceppi},
		\citenamefont {Brient}, \citenamefont {Zelinka},\ and\ \citenamefont
		{Hartmann}}]{Ceppi}%
	\BibitemOpen
	\bibfield  {author} {\bibinfo {author} {\bibfnamefont {P.}~\bibnamefont
			{Ceppi}}, \bibinfo {author} {\bibfnamefont {F.}~\bibnamefont {Brient}},
		\bibinfo {author} {\bibfnamefont {M.~D.}\ \bibnamefont {Zelinka}},\ and\
		\bibinfo {author} {\bibfnamefont {D.~L.}\ \bibnamefont {Hartmann}},\
	}\bibfield  {title} {\bibinfo {title} {Cloud feedback mechanisms and their
			representation in global climate models},\ }\href@noop {} {\bibfield
		{journal} {\bibinfo  {journal} {Wiley Interdisciplinary Reviews: Climate
				Change}\ }\textbf {\bibinfo {volume} {8(4)}} (\bibinfo {year}
		{2017})}\BibitemShut {NoStop}%
	\bibitem [{\citenamefont {Khouider}\ and\ \citenamefont
		{Bihlo}(2019)}]{Khouider}%
	\BibitemOpen
	\bibfield  {author} {\bibinfo {author} {\bibfnamefont {B.}~\bibnamefont
			{Khouider}}\ and\ \bibinfo {author} {\bibfnamefont {A.}~\bibnamefont
			{Bihlo}},\ }\bibfield  {title} {\bibinfo {title} {A new stochastic model for
			the boundary layer clouds and stratocumulus phase transition regimes: Open
			cells, closed cells, and convective rolls.},\ }\href@noop {} {\bibfield
		{journal} {\bibinfo  {journal} {Journal of Geophysical Research:
				Atmospheres,}\ }\textbf {\bibinfo {volume} {124}},\ \bibinfo {pages}
		{367–386} (\bibinfo {year} {2019})}\BibitemShut {NoStop}%
	\bibitem [{\citenamefont {I.~Tobin~and}\ and\ \citenamefont
		{Roca}(2012)}]{Tobin}%
	\BibitemOpen
	\bibfield  {author} {\bibinfo {author} {\bibfnamefont {S.~B.}\ \bibnamefont
			{I.~Tobin~and}}\ and\ \bibinfo {author} {\bibfnamefont {R.}~\bibnamefont
			{Roca}},\ }\bibfield  {title} {\bibinfo {title} {Observational evidence for
			relationships between the degree of aggregation of deep convection, water
			vapor, surface fluxes, and radiation},\ }\href@noop {} {\bibfield  {journal}
		{\bibinfo  {journal} {Journal of Climate}\ }\textbf {\bibinfo {volume}
			{25}},\ \bibinfo {pages} {6885–6904} (\bibinfo {year} {2012})}\BibitemShut
	{NoStop}%
	\bibitem [{\citenamefont {Robert A.~Houze}(1998)}]{Houze}%
	\BibitemOpen
	\bibfield  {author} {\bibinfo {author} {\bibfnamefont {J.}~\bibnamefont
			{Robert A.~Houze}},\ }\href@noop {} {\emph {\bibinfo {title} {Cloud
				Dynamics}}}\ (\bibinfo  {publisher} {{Elsevier B.V.}},\ \bibinfo {address}
	{Oxford, UK},\ \bibinfo {year} {1998})\BibitemShut {NoStop}%
	\bibitem [{\citenamefont {Vogel}\ \emph {et~al.}(2019)\citenamefont {Vogel},
		\citenamefont {Nuijens},\ and\ \citenamefont {Stevens}}]{Vogel}%
	\BibitemOpen
	\bibfield  {author} {\bibinfo {author} {\bibfnamefont {R.}~\bibnamefont
			{Vogel}}, \bibinfo {author} {\bibfnamefont {L.}~\bibnamefont {Nuijens}},\
		and\ \bibinfo {author} {\bibfnamefont {B.}~\bibnamefont {Stevens}},\
	}\bibfield  {title} {\bibinfo {title} {Influence of deepening and mesoscale
			organization of shallow convection on stratiform cloudiness in the downstream
			trades},\ }\href {https://doi.org/10.1007/s40641-019-00126-x} {\bibfield
		{journal} {\bibinfo  {journal} {Quarterly Journal of the Royal Meteorological
				Society}\ }\textbf {\bibinfo {volume} {146}},\ \bibinfo {pages} {174–185}
		(\bibinfo {year} {2019})}\BibitemShut {NoStop}%
	\bibitem [{\citenamefont {Wood}\ and\ \citenamefont {Taylor}(2006)}]{Taylor}%
	\BibitemOpen
	\bibfield  {author} {\bibinfo {author} {\bibfnamefont {C.~R.}\ \bibnamefont
			{Wood}}\ and\ \bibinfo {author} {\bibfnamefont {J.~P.}\ \bibnamefont
			{Taylor}},\ }\bibfield  {title} {\bibinfo {title} {Liquid water path
			variability in unbroken marine stratocumulus cloud},\ }\href@noop {}
	{\bibfield  {journal} {\bibinfo  {journal} {Quarterly Journal of the Royal
				Meteorological}\ }\textbf {\bibinfo {volume} {127(578)}} (\bibinfo {year}
		{2006})}\BibitemShut {NoStop}%
	\bibitem [{\citenamefont {Schneider}\ \emph {et~al.}(2019)\citenamefont
		{Schneider}, \citenamefont {Colleen},\ and\ \citenamefont
		{Pressel}}]{Colleen}%
	\BibitemOpen
	\bibfield  {author} {\bibinfo {author} {\bibfnamefont {T.}~\bibnamefont
			{Schneider}}, \bibinfo {author} {\bibfnamefont {M.~K.}\ \bibnamefont
			{Colleen}},\ and\ \bibinfo {author} {\bibfnamefont {K.~G.}\ \bibnamefont
			{Pressel}},\ }\bibfield  {title} {\bibinfo {title} {Possible climate
			transitions from breakup of stratocumulus decks under greenhouse warming},\
	}\href {https://doi.org/10.1038/s41561-019-0310-1} {\bibfield  {journal}
		{\bibinfo  {journal} {Nature Geoscience}\ }\textbf {\bibinfo {volume}
			{12(3)}},\ \bibinfo {pages} {164} (\bibinfo {year} {2019})}\BibitemShut
	{NoStop}%
	\bibitem [{\citenamefont {Wood}(2012)}]{Wood}%
	\BibitemOpen
	\bibfield  {author} {\bibinfo {author} {\bibfnamefont {R.}~\bibnamefont
			{Wood}},\ }\bibfield  {title} {\bibinfo {title} {Review: Stratocumulus
			clouds},\ }\href {https://doi.org/10.1175/MWR-D-11-00121.1} {\bibfield
		{journal} {\bibinfo  {journal} {Monthly weather review}\ }\textbf {\bibinfo
			{volume} {140}},\ \bibinfo {pages} {2373} (\bibinfo {year}
		{2012})}\BibitemShut {NoStop}%
	\bibitem [{\citenamefont {Noteboom}(2006)}]{Noteboom}%
	\BibitemOpen
	\bibfield  {author} {\bibinfo {author} {\bibfnamefont {S.}~\bibnamefont
			{Noteboom}},\ }\href@noop {} {\emph {\bibinfo {title} {Open Cell Convection
				and Closed Cell Convection}}}\ (\bibinfo  {publisher} {De Bilt: KNMI},\
	\bibinfo {year} {2006})\BibitemShut {NoStop}%
	\bibitem [{\citenamefont {Agee}\ \emph {et~al.}(1973)\citenamefont {Agee},
		\citenamefont {Chen},\ and\ \citenamefont {Dowell}}]{Agee}%
	\BibitemOpen
	\bibfield  {author} {\bibinfo {author} {\bibfnamefont {E.~M.}\ \bibnamefont
			{Agee}}, \bibinfo {author} {\bibfnamefont {T.~S.}\ \bibnamefont {Chen}},\
		and\ \bibinfo {author} {\bibfnamefont {K.~E.}\ \bibnamefont {Dowell}},\
	}\bibfield  {title} {\bibinfo {title} {A review of mesoscale cellular
			convection},\ }\href@noop {} {\bibfield  {journal} {\bibinfo  {journal}
			{Bulletin of the American Meteorological Society}\ }\textbf {\bibinfo
			{volume} {54}},\ \bibinfo {pages} {1004–1012} (\bibinfo {year}
		{1973})}\BibitemShut {NoStop}%
	\bibitem [{\citenamefont {Bretherton}\ and\ \citenamefont
		{Blossey}(2017)}]{Blossey}%
	\BibitemOpen
	\bibfield  {author} {\bibinfo {author} {\bibfnamefont {C.~S.}\ \bibnamefont
			{Bretherton}}\ and\ \bibinfo {author} {\bibfnamefont {P.~N.}\ \bibnamefont
			{Blossey}},\ }\bibfield  {title} {\bibinfo {title} {Understanding mesoscale
			aggregation of shallow cumulus convection using large‐eddy simulation},\
	}\href {https://doi.org/10.1002/2017MS000981} {\bibfield  {journal} {\bibinfo
			{journal} {Nature Geoscience}\ }\textbf {\bibinfo {volume} {9(8)}},\
		\bibinfo {pages} {2798} (\bibinfo {year} {2017})}\BibitemShut {NoStop}%
	\bibitem [{\citenamefont {Hottovy}\ and\ \citenamefont
		{Stechmann}(2015)}]{Hottovy}%
	\BibitemOpen
	\bibfield  {author} {\bibinfo {author} {\bibfnamefont {S.}~\bibnamefont
			{Hottovy}}\ and\ \bibinfo {author} {\bibfnamefont {S.~N.}\ \bibnamefont
			{Stechmann}},\ }\bibfield  {title} {\bibinfo {title} {A spatiotemporal
			stochastic model for tropical precipitation and water vapor dynamics},\
	}\href {https://doi.org/10.1175/JAS-D-15-0119.1} {\bibfield  {journal}
		{\bibinfo  {journal} {Journal of the Atmosferic Sciences}\ }\textbf {\bibinfo
			{volume} {72}},\ \bibinfo {pages} {4721} (\bibinfo {year}
		{2015})}\BibitemShut {NoStop}%
	\bibitem [{Sup()}]{Supplemental}%
	\BibitemOpen
	\href@noop {} {}\bibinfo {note} {See Supplemental Material at [URL will be
		inserted by publisher] for more details on the numerical solutions of the
		models presented, specifically the parameters values used in each cloud
		regime.}\BibitemShut {Stop}%
	\bibitem [{\citenamefont {McCoy}\ \emph {et~al.}(2017)\citenamefont {McCoy},
		\citenamefont {Wood},\ and\ \citenamefont {Fletcher}}]{McCoy}%
	\BibitemOpen
	\bibfield  {author} {\bibinfo {author} {\bibfnamefont {I.~L.}\ \bibnamefont
			{McCoy}}, \bibinfo {author} {\bibfnamefont {R.}~\bibnamefont {Wood}},\ and\
		\bibinfo {author} {\bibfnamefont {J.~K.}\ \bibnamefont {Fletcher}},\
	}\bibfield  {title} {\bibinfo {title} {Identifying meteorological controls on
			open and closed mesoscale cellular convection associated with marine cold air
			outbreaks},\ }\href@noop {} {\bibfield  {journal} {\bibinfo  {journal}
			{Journal of Geophysical Research: Atmospheres}\ }\textbf {\bibinfo {volume}
			{122(21)}} (\bibinfo {year} {2017})}\BibitemShut {NoStop}%
	\bibitem [{\citenamefont {Feingold}\ \emph {et~al.}(2005)\citenamefont
		{Feingold}, \citenamefont {Koren}, \citenamefont {Yamaguchi},\ and\
		\citenamefont {Kazil}}]{Koren}%
	\BibitemOpen
	\bibfield  {author} {\bibinfo {author} {\bibfnamefont {G.}~\bibnamefont
			{Feingold}}, \bibinfo {author} {\bibfnamefont {I.}~\bibnamefont {Koren}},
		\bibinfo {author} {\bibfnamefont {T.}~\bibnamefont {Yamaguchi}},\ and\
		\bibinfo {author} {\bibfnamefont {J.}~\bibnamefont {Kazil}},\ }\bibfield
	{title} {\bibinfo {title} {On the reversibility of transitions between closed
			and open cellular convection},\ }\href
	{https://doi.org/10.5194/acp-15-7351-2015} {\bibfield  {journal} {\bibinfo
			{journal} {Atmospheric Chemistry Physics}\ }\textbf {\bibinfo {volume}
			{15}},\ \bibinfo {pages} {7351–7367} (\bibinfo {year} {2005})}\BibitemShut
	{NoStop}%
	\bibitem [{\citenamefont {Yamaguchi}\ and\ \citenamefont
		{Feingold}(2015)}]{Yamaguchi}%
	\BibitemOpen
	\bibfield  {author} {\bibinfo {author} {\bibfnamefont {T.}~\bibnamefont
			{Yamaguchi}}\ and\ \bibinfo {author} {\bibfnamefont {G.}~\bibnamefont
			{Feingold}},\ }\bibfield  {title} {\bibinfo {title} {On the relationship
			between open cellular convective cloud patterns and the spatial distribution
			of precipitation.},\ }\href@noop {} {\bibfield  {journal} {\bibinfo
			{journal} {Atmospheric Chemistry and Physics}\ }\textbf {\bibinfo {volume}
			{15(3)}},\ \bibinfo {pages} {237–1251} (\bibinfo {year}
		{2015})}\BibitemShut {NoStop}%
	\bibitem [{\citenamefont {Peters}\ \emph {et~al.}(2009)\citenamefont {Peters},
		\citenamefont {Neelin},\ and\ \citenamefont {Nesbitt}}]{Nesbitt}%
	\BibitemOpen
	\bibfield  {author} {\bibinfo {author} {\bibfnamefont {O.}~\bibnamefont
			{Peters}}, \bibinfo {author} {\bibfnamefont {J.~D.}\ \bibnamefont {Neelin}},\
		and\ \bibinfo {author} {\bibfnamefont {S.~W.}\ \bibnamefont {Nesbitt}},\
	}\bibfield  {title} {\bibinfo {title} {Mesoscale convective systems and
			critical clusters},\ }\href {https://doi.org/10.1175/2008JAS2761.1}
	{\bibfield  {journal} {\bibinfo  {journal} {Journal of the Atmosferic
				Sciences}\ }\textbf {\bibinfo {volume} {66}},\ \bibinfo {pages} {2913}
		(\bibinfo {year} {2009})}\BibitemShut {NoStop}%
	\bibitem [{\citenamefont {Glassmeiera}\ and\ \citenamefont
		{Feingold}(2017)}]{Feingold}%
	\BibitemOpen
	\bibfield  {author} {\bibinfo {author} {\bibfnamefont {F.}~\bibnamefont
			{Glassmeiera}}\ and\ \bibinfo {author} {\bibfnamefont {G.}~\bibnamefont
			{Feingold}},\ }\bibfield  {title} {\bibinfo {title} {Network approach to
			patterns in stratocumulus clouds},\ }\href
	{https://doi.org/10.1073/pnas.1706495114} {\bibfield  {journal} {\bibinfo
			{journal} {PNAS}\ }\textbf {\bibinfo {volume} {114}},\ \bibinfo {pages}
		{10578} (\bibinfo {year} {2017})}\BibitemShut {NoStop}%
	\bibitem [{\citenamefont {Bak}\ \emph {et~al.}(1988)\citenamefont {Bak},
		\citenamefont {Tang},\ and\ \citenamefont {Wiesenfeld}}]{Bak}%
	\BibitemOpen
	\bibfield  {author} {\bibinfo {author} {\bibfnamefont {P.}~\bibnamefont
			{Bak}}, \bibinfo {author} {\bibfnamefont {C.}~\bibnamefont {Tang}},\ and\
		\bibinfo {author} {\bibfnamefont {K.}~\bibnamefont {Wiesenfeld}},\ }\bibfield
	{title} {\bibinfo {title} {Self-organized criticality},\ }\href@noop {}
	{\bibfield  {journal} {\bibinfo  {journal} {Physical Review A}\ }\textbf
		{\bibinfo {volume} {38(1)}},\ \bibinfo {pages} {364} (\bibinfo {year}
		{1988})}\BibitemShut {NoStop}%
	\bibitem [{\citenamefont {Peters}\ and\ \citenamefont {Neelin}(2006)}]{Peters}%
	\BibitemOpen
	\bibfield  {author} {\bibinfo {author} {\bibfnamefont {O.}~\bibnamefont
			{Peters}}\ and\ \bibinfo {author} {\bibfnamefont {J.~D.}\ \bibnamefont
			{Neelin}},\ }\bibfield  {title} {\bibinfo {title} {Critical phenomena in
			atmospheric precipitation},\ }\href {https://doi.org/10.1038/nphys314}
	{\bibfield  {journal} {\bibinfo  {journal} {Nature Physics}\ }\textbf
		{\bibinfo {volume} {2}},\ \bibinfo {pages} {393} (\bibinfo {year}
		{2006})}\BibitemShut {NoStop}%
	\bibitem [{\citenamefont {Hottovy}\ and\ \citenamefont
		{Stechmann}(2016)}]{Stechmann}%
	\BibitemOpen
	\bibfield  {author} {\bibinfo {author} {\bibfnamefont {S.}~\bibnamefont
			{Hottovy}}\ and\ \bibinfo {author} {\bibfnamefont {S.~N.}\ \bibnamefont
			{Stechmann}},\ }\bibfield  {title} {\bibinfo {title} {Cloud regimes as phase
			transitions},\ }\href {https://doi.org/10.1002/2016GL069396} {\bibfield
		{journal} {\bibinfo  {journal} {Geophysical Research Letters}\ }\textbf
		{\bibinfo {volume} {43}},\ \bibinfo {pages} {6579–6587} (\bibinfo {year}
		{2016})}\BibitemShut {NoStop}%
	\bibitem [{\citenamefont {B.}\ \emph {et~al.}(2005)\citenamefont {B.},
		\citenamefont {Vali}, \citenamefont {Comstock}, \citenamefont {Wood},
		\citenamefont {van Zanten}, \citenamefont {P.H}, \citenamefont {Austin},\
		and\ \citenamefont {Lenschow}}]{Stevens}%
	\BibitemOpen
	\bibfield  {author} {\bibinfo {author} {\bibfnamefont {S.}~\bibnamefont
			{B.}}, \bibinfo {author} {\bibfnamefont {G.}~\bibnamefont {Vali}}, \bibinfo
		{author} {\bibfnamefont {K.}~\bibnamefont {Comstock}}, \bibinfo {author}
		{\bibfnamefont {R.}~\bibnamefont {Wood}}, \bibinfo {author} {\bibfnamefont
			{M.}~\bibnamefont {van Zanten}}, \bibinfo {author} {\bibnamefont {P.H}},
		\bibinfo {author} {\bibfnamefont {C.~B.}\ \bibnamefont {Austin}},\ and\
		\bibinfo {author} {\bibfnamefont {D.}~\bibnamefont {Lenschow}},\ }\bibfield
	{title} {\bibinfo {title} {Pockets of open cells ans drizzle in marine
			stratocumulus},\ }\href@noop {} {\bibfield  {journal} {\bibinfo  {journal}
			{Bulletin of the American Meteorological Society}\ }\textbf {\bibinfo
			{volume} {86}},\ \bibinfo {pages} {51–58} (\bibinfo {year}
		{2005})}\BibitemShut {NoStop}%
	\bibitem [{\citenamefont {Ahmed}\ and\ \citenamefont
		{Schumacher}(2015)}]{Ahmed}%
	\BibitemOpen
	\bibfield  {author} {\bibinfo {author} {\bibfnamefont {F.}~\bibnamefont
			{Ahmed}}\ and\ \bibinfo {author} {\bibfnamefont {C.}~\bibnamefont
			{Schumacher}},\ }\bibfield  {title} {\bibinfo {title} {Convective and
			stratiform components of the precipitation‐moisture relationship},\
	}\href@noop {} {\bibfield  {journal} {\bibinfo  {journal} {Geophysical
				Research Letters}\ }\textbf {\bibinfo {volume} {42 (23)}} (\bibinfo {year}
		{2015})}\BibitemShut {NoStop}%
	\bibitem [{\citenamefont {Craig}\ and\ \citenamefont
		{Mack}(2013)}]{Craig_2013}%
	\BibitemOpen
	\bibfield  {author} {\bibinfo {author} {\bibfnamefont {G.~C.}\ \bibnamefont
			{Craig}}\ and\ \bibinfo {author} {\bibfnamefont {J.~M.}\ \bibnamefont
			{Mack}},\ }\bibfield  {title} {\bibinfo {title} {A coarsening model for
			self-organization of tropical convection},\ }\href
	{https://doi.org/10.1002/jgrd.50674} {\bibfield  {journal} {\bibinfo
			{journal} {Journal of Geophysical Research: Atmospheres}\ }\textbf {\bibinfo
			{volume} {118}},\ \bibinfo {pages} {8761} (\bibinfo {year}
		{2013})}\BibitemShut {NoStop}%
	\bibitem [{\citenamefont {Allen}\ and\ \citenamefont
		{Cahn}(1979)}]{Allen_1979}%
	\BibitemOpen
	\bibfield  {author} {\bibinfo {author} {\bibfnamefont {S.~M.}\ \bibnamefont
			{Allen}}\ and\ \bibinfo {author} {\bibfnamefont {J.~W.}\ \bibnamefont
			{Cahn}},\ }\bibfield  {title} {\bibinfo {title} {A microscopic theory for
			antiphase boundary motion and its application to antiphase domain
			coarsening},\ }\href
	{https://doi.org/https://doi.org/10.1016/0001-6160(79)90196-2} {\bibfield
		{journal} {\bibinfo  {journal} {Acta Metallurgica}\ }\textbf {\bibinfo
			{volume} {27}},\ \bibinfo {pages} {1085 } (\bibinfo {year}
		{1979})}\BibitemShut {NoStop}%
	\bibitem [{\citenamefont {Jordi Garc\'ia-Ojalvo}(1999)}]{Sancho_1999}%
	\BibitemOpen
	\bibfield  {author} {\bibinfo {author} {\bibfnamefont {J.~M.~S.}\
			\bibnamefont {Jordi Garc\'ia-Ojalvo}},\ }\href@noop {} {\emph {\bibinfo
			{title} {Noise in Spatially Extended Systems}}},\ \bibinfo {edition} {1st}\
	ed.,\ Institute for Nonlinear Science\ (\bibinfo  {publisher}
	{Springer-Verlag New York},\ \bibinfo {year} {1999})\BibitemShut {NoStop}%
	\bibitem [{\citenamefont {Loecher}(2003)}]{Leocher_2003}%
	\BibitemOpen
	\bibfield  {author} {\bibinfo {author} {\bibfnamefont {M.}~\bibnamefont
			{Loecher}},\ }\href@noop {} {\emph {\bibinfo {title} {Noise sustained
				patterns : fluctuations and nonlinearities}}},\ \bibinfo {edition} {1st}\
	ed.,\ \bibinfo {series} {World Scientific Lecture Notes in Physics},
	Vol.~\bibinfo {volume} {70}\ (\bibinfo  {publisher} {World Scientific
		Singapour},\ \bibinfo {year} {2003})\BibitemShut {NoStop}%
	\bibitem [{\citenamefont {Betts}\ and\ \citenamefont {Miller}(1986)}]{Betts}%
	\BibitemOpen
	\bibfield  {author} {\bibinfo {author} {\bibfnamefont {A.~K.}\ \bibnamefont
			{Betts}}\ and\ \bibinfo {author} {\bibfnamefont {M.~J.}\ \bibnamefont
			{Miller}},\ }\bibfield  {title} {\bibinfo {title} {A new convective
			adjustment scheme. part ii: Single column tests using gate wave, bomex, atex
			and arctic air-mass data sets.},\ }\href@noop {} {\bibfield  {journal}
		{\bibinfo  {journal} {Journal of Geophysical Research: Atmospheres}\ }\textbf
		{\bibinfo {volume} {112(473)}},\ \bibinfo {pages} {693–709} (\bibinfo
		{year} {1986})}\BibitemShut {NoStop}%
	\bibitem [{\citenamefont {Majda}\ and\ \citenamefont {Grote}(2009)}]{Majda}%
	\BibitemOpen
	\bibfield  {author} {\bibinfo {author} {\bibfnamefont {A.~J.}\ \bibnamefont
			{Majda}}\ and\ \bibinfo {author} {\bibfnamefont {M.~J.}\ \bibnamefont
			{Grote}},\ }\bibfield  {title} {\bibinfo {title} {Mathematical test models
			for superparametrization in anisotropic turbulence.},\ }\href
	{https://doi.org/10.1073/pnas.0901383106} {\bibfield  {journal} {\bibinfo
			{journal} {Proceedings of the National Academy of Sciences}\ }\textbf
		{\bibinfo {volume} {106(14)}},\ \bibinfo {pages} {5470–5474} (\bibinfo
		{year} {2009})}\BibitemShut {NoStop}%
	\bibitem [{\citenamefont {Gurevich}(2017)}]{Gurevich}%
	\BibitemOpen
	\bibfield  {author} {\bibinfo {author} {\bibfnamefont {S.~V.}\ \bibnamefont
			{Gurevich}},\ }\href@noop {} {\emph {\bibinfo {title} {Numerical methods for
				complex systems II}}}\ (\bibinfo  {publisher} {Westfälische
		Wilhelms-Universität},\ \bibinfo {year} {2017})\BibitemShut {NoStop}%
	\bibitem [{\citenamefont {Stechmann}\ and\ \citenamefont
		{Neelin}(2011)}]{Neelin}%
	\BibitemOpen
	\bibfield  {author} {\bibinfo {author} {\bibfnamefont {S.~N.}\ \bibnamefont
			{Stechmann}}\ and\ \bibinfo {author} {\bibfnamefont {J.~D.}\ \bibnamefont
			{Neelin}},\ }\bibfield  {title} {\bibinfo {title} {A stochastic model for the
			transition to stong convection},\ }\href@noop {} {\bibfield  {journal}
		{\bibinfo  {journal} {Journal of the Atmospheric Sciences}\ }\textbf
		{\bibinfo {volume} {68(12)}},\ \bibinfo {pages} {2955} (\bibinfo {year}
		{2011})}\BibitemShut {NoStop}%
	\bibitem [{\citenamefont {Holloway}\ and\ \citenamefont
		{Neelin}(2009)}]{Holloway}%
	\BibitemOpen
	\bibfield  {author} {\bibinfo {author} {\bibfnamefont {C.~E.}\ \bibnamefont
			{Holloway}}\ and\ \bibinfo {author} {\bibfnamefont {J.~D.}\ \bibnamefont
			{Neelin}},\ }\bibfield  {title} {\bibinfo {title} {Moisture vertical
			structure, column water vapor, and tropical deep convection},\ }\href@noop {}
	{\bibfield  {journal} {\bibinfo  {journal} {Journal of Atmospheric Sciences}\
		}\textbf {\bibinfo {volume} {66}},\ \bibinfo {pages} {1665} (\bibinfo {year}
		{2009})}\BibitemShut {NoStop}%
	\bibitem [{\citenamefont {Bretherton}\ \emph {et~al.}(2004)\citenamefont
		{Bretherton}, \citenamefont {Peters},\ and\ \citenamefont
		{Back}}]{Bretherton}%
	\BibitemOpen
	\bibfield  {author} {\bibinfo {author} {\bibfnamefont {C.~S.}\ \bibnamefont
			{Bretherton}}, \bibinfo {author} {\bibfnamefont {M.~E.}\ \bibnamefont
			{Peters}},\ and\ \bibinfo {author} {\bibfnamefont {L.~E.}\ \bibnamefont
			{Back}},\ }\bibfield  {title} {\bibinfo {title} {Relationships between water
			vapor path and precipitation over the tropical oceans},\ }\href@noop {}
	{\bibfield  {journal} {\bibinfo  {journal} {Journal of Climate}\ }\textbf
		{\bibinfo {volume} {17}},\ \bibinfo {pages} {1517–1528} (\bibinfo {year}
		{2004})}\BibitemShut {NoStop}%
	\bibitem [{\citenamefont {Yano}\ \emph {et~al.}(2012)\citenamefont {Yano},
		\citenamefont {Liu},\ and\ \citenamefont {Moncrieff}}]{Yano}%
	\BibitemOpen
	\bibfield  {author} {\bibinfo {author} {\bibfnamefont {J.-I.}\ \bibnamefont
			{Yano}}, \bibinfo {author} {\bibfnamefont {C.}~\bibnamefont {Liu}},\ and\
		\bibinfo {author} {\bibfnamefont {M.~W.}\ \bibnamefont {Moncrieff}},\
	}\bibfield  {title} {\bibinfo {title} {Self-organized criticality and
			homeostasis in atmospheric convective organization},\ }\href@noop {}
	{\bibfield  {journal} {\bibinfo  {journal} {Journal of the Atmospheric
				Sciences}\ }\textbf {\bibinfo {volume} {69}},\ \bibinfo {pages} {3449}
		(\bibinfo {year} {2012})}\BibitemShut {NoStop}%
	\bibitem [{\citenamefont {Lebsock}\ \emph {et~al.}(2017)\citenamefont
		{Lebsock}, \citenamefont {L’Ecuyer},\ and\ \citenamefont
		{Pincus}}]{Lebsock}%
	\BibitemOpen
	\bibfield  {author} {\bibinfo {author} {\bibfnamefont {M.~D.}\ \bibnamefont
			{Lebsock}}, \bibinfo {author} {\bibfnamefont {T.~S.}\ \bibnamefont
			{L’Ecuyer}},\ and\ \bibinfo {author} {\bibfnamefont {R.}~\bibnamefont
			{Pincus}},\ }\bibfield  {title} {\bibinfo {title} {An observational view of
			relationships between moisture aggregation, cloud, and radiative heating
			profiles.},\ }\href@noop {} {\bibfield  {journal} {\bibinfo  {journal}
			{Surveys in Geophysics}\ }\textbf {\bibinfo {volume} {38}},\ \bibinfo {pages}
		{1237–1254} (\bibinfo {year} {2017})}\BibitemShut {NoStop}%
	\bibitem [{\citenamefont {Holloway}\ and\ \citenamefont
		{Neelin}(2010)}]{HollowayA}%
	\BibitemOpen
	\bibfield  {author} {\bibinfo {author} {\bibfnamefont {C.~E.}\ \bibnamefont
			{Holloway}}\ and\ \bibinfo {author} {\bibfnamefont {J.~D.}\ \bibnamefont
			{Neelin}},\ }\bibfield  {title} {\bibinfo {title} {Temporal relations of
			column water vapor and tropical precipitation.},\ }\href@noop {} {\bibfield
		{journal} {\bibinfo  {journal} {Journal of the Atmospheric Sciences}\
		}\textbf {\bibinfo {volume} {67}},\ \bibinfo {pages} {1091–1105} (\bibinfo
		{year} {2010})}\BibitemShut {NoStop}%
	\bibitem [{\citenamefont {Binney}\ \emph {et~al.}(1992)\citenamefont {Binney},
		\citenamefont {Dowrick}, \citenamefont {Fisher},\ and\ \citenamefont
		{Newman}}]{Binney}%
	\BibitemOpen
	\bibfield  {author} {\bibinfo {author} {\bibfnamefont {J.~J.}\ \bibnamefont
			{Binney}}, \bibinfo {author} {\bibfnamefont {N.~J.}\ \bibnamefont {Dowrick}},
		\bibinfo {author} {\bibfnamefont {A.~J.}\ \bibnamefont {Fisher}},\ and\
		\bibinfo {author} {\bibfnamefont {M.~E.~J.}\ \bibnamefont {Newman}},\
	}\bibinfo {title} {The theory of critical phenomena: An introduction to the
		renormalization group}\ (\bibinfo  {publisher} {Oxford Science
		Publications},\ \bibinfo {year} {1992})\ Chap.~\bibinfo {chapter}
	{7}\BibitemShut {NoStop}%
	\bibitem [{\citenamefont {Komin}\ \emph {et~al.}(2010)\citenamefont {Komin},
		\citenamefont {Lacasa},\ and\ \citenamefont {Toral}}]{Komin}%
	\BibitemOpen
	\bibfield  {author} {\bibinfo {author} {\bibfnamefont {N.}~\bibnamefont
			{Komin}}, \bibinfo {author} {\bibfnamefont {L.}~\bibnamefont {Lacasa}},\ and\
		\bibinfo {author} {\bibfnamefont {R.}~\bibnamefont {Toral}},\ }\bibfield
	{title} {\bibinfo {title} {Critical behavior of a ginzburg–landau model
			with additive quenched noise.},\ }\href
	{https://doi.org/10.1088/1742-5468/2010/12/p12008} {\bibfield  {journal}
		{\bibinfo  {journal} {Journal of Statistical Mechanics: Theory and
				Experiment}\ }\textbf {\bibinfo {volume} {12}},\ \bibinfo {pages} {P12008}
		(\bibinfo {year} {2010})}\BibitemShut {NoStop}%
	\bibitem [{\citenamefont {Hern{\'a}ndez-Machado}\ \emph
		{et~al.}(1993)\citenamefont {Hern{\'a}ndez-Machado}, \citenamefont
		{Ram{\'i}rez-Piscina},\ and\ \citenamefont {Sancho}}]{Machado}%
	\BibitemOpen
	\bibfield  {author} {\bibinfo {author} {\bibfnamefont {A.}~\bibnamefont
			{Hern{\'a}ndez-Machado}}, \bibinfo {author} {\bibfnamefont {L.}~\bibnamefont
			{Ram{\'i}rez-Piscina}},\ and\ \bibinfo {author} {\bibfnamefont {J.~M.}\
			\bibnamefont {Sancho}},\ }\bibfield  {title} {\bibinfo {title}
		{Multiplicative noise in domain growth: Stochastic ginzburg-landau
			equations},\ }\href {https://doi.org/10.1007/978-1-4615-2852-4_37} {\bibfield
		{journal} {\bibinfo  {journal} {Physical Review B}\ }\textbf {\bibinfo
			{volume} {48}},\ \bibinfo {pages} {125} (\bibinfo {year} {1993})}\BibitemShut
	{NoStop}%
	\bibitem [{\citenamefont {Yeomans}(1992)}]{Yeomans}%
	\BibitemOpen
	\bibfield  {author} {\bibinfo {author} {\bibfnamefont {J.~M.}\ \bibnamefont
			{Yeomans}},\ }\href@noop {} {\emph {\bibinfo {title} {Statistical Mechanics
				of Phase Transitions}}}\ (\bibinfo  {publisher} {Oxford Science
		Publications},\ \bibinfo {year} {1992})\BibitemShut {NoStop}%
	\bibitem [{\citenamefont {Watkins}\ \emph {et~al.}(2016)\citenamefont
		{Watkins}, \citenamefont {Pruessner}, \citenamefont {Chapman}, \citenamefont
		{Crosby},\ and\ \citenamefont {Jensen}}]{Watkins}%
	\BibitemOpen
	\bibfield  {author} {\bibinfo {author} {\bibfnamefont {N.~W.}\ \bibnamefont
			{Watkins}}, \bibinfo {author} {\bibfnamefont {G.}~\bibnamefont {Pruessner}},
		\bibinfo {author} {\bibfnamefont {S.~C.}\ \bibnamefont {Chapman}}, \bibinfo
		{author} {\bibfnamefont {N.~B.}\ \bibnamefont {Crosby}},\ and\ \bibinfo
		{author} {\bibfnamefont {H.~J.}\ \bibnamefont {Jensen}},\ }\bibfield  {title}
	{\bibinfo {title} {25 years of self-organized criticality: Concepts and
			controversies},\ }\href@noop {} {\bibfield  {journal} {\bibinfo  {journal}
			{Space Science Reviews}\ }\textbf {\bibinfo {volume} {198}},\ \bibinfo
		{pages} {3} (\bibinfo {year} {2016})}\BibitemShut {NoStop}%
	\bibitem [{\citenamefont {Ryser}\ \emph {et~al.}(2012)\citenamefont {Ryser},
		\citenamefont {Nigam},\ and\ \citenamefont {Tupper}}]{Ryser_2012}%
	\BibitemOpen
	\bibfield  {author} {\bibinfo {author} {\bibfnamefont {M.~D.}\ \bibnamefont
			{Ryser}}, \bibinfo {author} {\bibfnamefont {N.}~\bibnamefont {Nigam}},\ and\
		\bibinfo {author} {\bibfnamefont {P.~F.}\ \bibnamefont {Tupper}},\ }\bibfield
	{title} {\bibinfo {title} {On the well-posedness of the stochastic
			allen–cahn equation in two dimensions},\ }\href
	{https://doi.org/https://doi.org/10.1016/j.jcp.2011.12.002} {\bibfield
		{journal} {\bibinfo  {journal} {Journal of Computational Physics}\ }\textbf
		{\bibinfo {volume} {231}},\ \bibinfo {pages} {2537 } (\bibinfo {year}
		{2012})}\BibitemShut {NoStop}%
	\bibitem [{\citenamefont {Krishnamurti}(1975)}]{Krishnamurti_1975}%
	\BibitemOpen
	\bibfield  {author} {\bibinfo {author} {\bibfnamefont {R.}~\bibnamefont
			{Krishnamurti}},\ }\bibfield  {title} {\bibinfo {title} {{On Cellular Cloud
				Patterns. Part 1: Mathematical Model}},\ }\href
	{https://doi.org/10.1175/1520-0469(1975)032<1353:OCCPM>2.0.CO;2} {\bibfield
		{journal} {\bibinfo  {journal} {Journal of the Atmospheric Sciences}\
		}\textbf {\bibinfo {volume} {32}},\ \bibinfo {pages} {1353} (\bibinfo {year}
		{1975})}\BibitemShut {NoStop}%
	\bibitem [{\citenamefont {Cross}(2006)}]{Croos3}%
	\BibitemOpen
	\bibfield  {author} {\bibinfo {author} {\bibfnamefont {M.}~\bibnamefont
			{Cross}},\ }\bibfield  {title} {\bibinfo {title} {Lecture 8 supplementary
			notes: Amplitude equations},\ }\href@noop {} {\bibfield  {journal} {\bibinfo
			{journal} {California Institute of Technology}\ } (\bibinfo {year}
		{2006})}\BibitemShut {NoStop}%
	\bibitem [{\citenamefont {van Hecke}\ \emph {et~al.}(1994)\citenamefont {van
			Hecke}, \citenamefont {Hohenberg},\ and\ \citenamefont {van
			Saarloos}}]{Hecke}%
	\BibitemOpen
	\bibfield  {author} {\bibinfo {author} {\bibfnamefont {M.}~\bibnamefont {van
				Hecke}}, \bibinfo {author} {\bibfnamefont {P.~C.}\ \bibnamefont
			{Hohenberg}},\ and\ \bibinfo {author} {\bibfnamefont {W.}~\bibnamefont {van
				Saarloos}},\ }\bibfield  {title} {\bibinfo {title} {Amplitude equations for
			pattern forming systems.},\ }in\ \href
	{https://doi.org/10.1016/b978-0-444-81591-0.50014-6} {\emph {\bibinfo
			{booktitle} {Fundamental Problems in Statistical Mechanics}}},\ Vol.\
	\bibinfo {volume} {VIII},\ \bibinfo {editor} {edited by\ \bibinfo {editor}
		{\bibfnamefont {H.}~\bibnamefont {van Beijeren}}\ and\ \bibinfo {editor}
		{\bibfnamefont {M.~H.}\ \bibnamefont {Ernst}}}\ (\bibinfo  {publisher} {North
		Holland},\ \bibinfo {year} {1994})\ p.\ \bibinfo {pages}
	{245–278}\BibitemShut {NoStop}%
	\bibitem [{\citenamefont {Swift}\ and\ \citenamefont
		{Hohenberg}(1977)}]{Swift_1977}%
	\BibitemOpen
	\bibfield  {author} {\bibinfo {author} {\bibfnamefont {J.}~\bibnamefont
			{Swift}}\ and\ \bibinfo {author} {\bibfnamefont {P.~C.}\ \bibnamefont
			{Hohenberg}},\ }\bibfield  {title} {\bibinfo {title} {Hydrodynamic
			fluctuations at the convective instability},\ }\href
	{https://doi.org/10.1103/PhysRevA.15.319} {\bibfield  {journal} {\bibinfo
			{journal} {Phys. Rev. A}\ }\textbf {\bibinfo {volume} {15}},\ \bibinfo
		{pages} {319} (\bibinfo {year} {1977})}\BibitemShut {NoStop}%
	\bibitem [{\citenamefont {Gao}(2017)}]{Gao_2017}%
	\BibitemOpen
	\bibfield  {author} {\bibinfo {author} {\bibfnamefont {P.}~\bibnamefont
			{Gao}},\ }\bibfield  {title} {\bibinfo {title} {The stochastic
			swift{\textendash}hohenberg equation},\ }\href
	{https://doi.org/10.1088/1361-6544/aa7e99} {\bibfield  {journal} {\bibinfo
			{journal} {Nonlinearity}\ }\textbf {\bibinfo {volume} {30}},\ \bibinfo
		{pages} {3516} (\bibinfo {year} {2017})}\BibitemShut {NoStop}%
	\bibitem [{\citenamefont {Garc\'{\i}a-Ojalvo}\ \emph
		{et~al.}(1993)\citenamefont {Garc\'{\i}a-Ojalvo}, \citenamefont
		{Hern\'andez-Machado},\ and\ \citenamefont {Sancho}}]{PhysRevLett.71.1542}%
	\BibitemOpen
	\bibfield  {author} {\bibinfo {author} {\bibfnamefont {J.}~\bibnamefont
			{Garc\'{\i}a-Ojalvo}}, \bibinfo {author} {\bibfnamefont {A.}~\bibnamefont
			{Hern\'andez-Machado}},\ and\ \bibinfo {author} {\bibfnamefont {J.~M.}\
			\bibnamefont {Sancho}},\ }\bibfield  {title} {\bibinfo {title} {Effects of
			external noise on the swift-hohenberg equation},\ }\href
	{https://doi.org/10.1103/PhysRevLett.71.1542} {\bibfield  {journal} {\bibinfo
			{journal} {Phys. Rev. Lett.}\ }\textbf {\bibinfo {volume} {71}},\ \bibinfo
		{pages} {1542} (\bibinfo {year} {1993})}\BibitemShut {NoStop}%
	\bibitem [{\citenamefont {Doelman}\ and\ \citenamefont
		{Schneider}(2001)}]{Doelman}%
	\BibitemOpen
	\bibfield  {author} {\bibinfo {author} {\bibfnamefont {A.}~\bibnamefont
			{Doelman}}\ and\ \bibinfo {author} {\bibfnamefont {G.}~\bibnamefont
			{Schneider}},\ }\href@noop {} {\bibinfo {title} {Lecture notes in the complex
			ginzburg-landau equation}} (\bibinfo {year} {2001}),\ \bibinfo {note}
	{universiteit van Amsterdam}\BibitemShut {NoStop}%
	\bibitem [{\citenamefont {Klepel}\ \emph {et~al.}(2013)\citenamefont {Klepel},
		\citenamefont {Blömker},\ and\ \citenamefont {Mohammed}}]{Klepel}%
	\BibitemOpen
	\bibfield  {author} {\bibinfo {author} {\bibfnamefont {K.}~\bibnamefont
			{Klepel}}, \bibinfo {author} {\bibfnamefont {D.}~\bibnamefont {Blömker}},\
		and\ \bibinfo {author} {\bibfnamefont {W.}~\bibnamefont {Mohammed}},\
	}\bibfield  {title} {\bibinfo {title} {Amplitude equation for the generalized
			swift hohenberg equation with noise},\ }\href
	{https://doi.org/10.1007/s00033-013-0371-8} {\bibfield  {journal} {\bibinfo
			{journal} {Zeitschrift für angewandte Mathematik und Physik}\ }\textbf
		{\bibinfo {volume} {65}},\ \bibinfo {pages} {1107–1126} (\bibinfo {year}
		{2013})}\BibitemShut {NoStop}%
	\bibitem [{\citenamefont {Saarloos}(1994)}]{Saarloos}%
	\BibitemOpen
	\bibfield  {author} {\bibinfo {author} {\bibfnamefont {W.}~\bibnamefont
			{Saarloos}},\ }\bibfield  {title} {\bibinfo {title} {The complex
			ginzburg–landau equation for beginners.},\ }\href@noop {} {\bibfield
		{journal} {\bibinfo  {journal} {Proceedings of the Santa Fe Workshop on
				Spatio-Temporal Patterns in Nonequilibrium Complex Systems.}\ ,\ \bibinfo
			{pages} {19–31}} (\bibinfo {year} {1994})}\BibitemShut {NoStop}%
	\bibitem [{\citenamefont {Pérez-Moreno}\ \emph {et~al.}(2014)\citenamefont
		{Pérez-Moreno}, \citenamefont {Chavarría},\ and\ \citenamefont
		{Ruiz-Chavarría}}]{Chavarria}%
	\BibitemOpen
	\bibfield  {author} {\bibinfo {author} {\bibfnamefont {S.~S.}\ \bibnamefont
			{Pérez-Moreno}}, \bibinfo {author} {\bibfnamefont {M.~S.~R.}\ \bibnamefont
			{Chavarría}},\ and\ \bibinfo {author} {\bibfnamefont {G.}~\bibnamefont
			{Ruiz-Chavarría}},\ }\bibfield  {title} {\bibinfo {title} {Numerical
			solution of the swift–hohenberg equation},\ }in\ \href@noop {} {\emph
		{\bibinfo {booktitle} {Experimental and Computational Fluid Mechanics.
				Environmental Science and Engineering}}},\ \bibinfo {editor} {edited by\
		\bibinfo {editor} {\bibfnamefont {J.}~\bibnamefont {Klapp}}\ and\ \bibinfo
		{editor} {\bibfnamefont {A.}~\bibnamefont {Medina}}}\ (\bibinfo  {publisher}
	{Springer International Publishing},\ \bibinfo {year} {2014})\ pp.\ \bibinfo
	{pages} {409--416}\BibitemShut {NoStop}%
\end{thebibliography}

\begin{thebibliography}{4}%
	\makeatletter
	\providecommand \@ifxundefined [1]{%
		\@ifx{#1\undefined}
	}%
	\providecommand \@ifnum [1]{%
		\ifnum #1\expandafter \@firstoftwo
		\else \expandafter \@secondoftwo
		\fi
	}%
	\providecommand \@ifx [1]{%
		\ifx #1\expandafter \@firstoftwo
		\else \expandafter \@secondoftwo
		\fi
	}%
	\providecommand \natexlab [1]{#1}%
	\providecommand \enquote  [1]{``#1''}%
	\providecommand \bibnamefont  [1]{#1}%
	\providecommand \bibfnamefont [1]{#1}%
	\providecommand \citenamefont [1]{#1}%
	\providecommand \href@noop [0]{\@secondoftwo}%
	\providecommand \href [0]{\begingroup \@sanitize@url \@href}%
	\providecommand \@href[1]{\@@startlink{#1}\@@href}%
	\providecommand \@@href[1]{\endgroup#1\@@endlink}%
	\providecommand \@sanitize@url [0]{\catcode `\\12\catcode `\$12\catcode
		`\&12\catcode `\#12\catcode `\^12\catcode `\_12\catcode `\%12\relax}%
	\providecommand \@@startlink[1]{}%
	\providecommand \@@endlink[0]{}%
	\providecommand \url  [0]{\begingroup\@sanitize@url \@url }%
	\providecommand \@url [1]{\endgroup\@href {#1}{\urlprefix }}%
	\providecommand \urlprefix  [0]{URL }%
	\providecommand \Eprint [0]{\href }%
	\providecommand \doibase [0]{https://doi.org/}%
	\providecommand \selectlanguage [0]{\@gobble}%
	\providecommand \bibinfo  [0]{\@secondoftwo}%
	\providecommand \bibfield  [0]{\@secondoftwo}%
	\providecommand \translation [1]{[#1]}%
	\providecommand \BibitemOpen [0]{}%
	\providecommand \bibitemStop [0]{}%
	\providecommand \bibitemNoStop [0]{.\EOS\space}%
	\providecommand \EOS [0]{\spacefactor3000\relax}%
	\providecommand \BibitemShut  [1]{\csname bibitem#1\endcsname}%
	\let\auto@bib@innerbib\@empty
	\bibitem [{\citenamefont {Hottovy}\ and\ \citenamefont
		{Stechmann}(2015)}]{Hottovy20}%
	\BibitemOpen
	\bibfield  {author} {\bibinfo {author} {\bibfnamefont {S.}~\bibnamefont
			{Hottovy}}\ and\ \bibinfo {author} {\bibfnamefont {S.~N.}\ \bibnamefont
			{Stechmann}},\ }\bibfield  {title} {\bibinfo {title} {A spatiotemporal
			stochastic model for tropical precipitation and water vapor dynamics},\
	}\href {https://doi.org/10.1175/JAS-D-15-0119.1} {\bibfield  {journal}
		{\bibinfo  {journal} {Journal of the Atmosferic Sciences}\ }\textbf {\bibinfo
			{volume} {72}},\ \bibinfo {pages} {4721} (\bibinfo {year}
		{2015})}\BibitemShut {NoStop}%
	\bibitem [{\citenamefont {Hottovy}\ and\ \citenamefont
		{Stechmann}(2016)}]{Stechmann20}%
	\BibitemOpen
	\bibfield  {author} {\bibinfo {author} {\bibfnamefont {S.}~\bibnamefont
			{Hottovy}}\ and\ \bibinfo {author} {\bibfnamefont {S.~N.}\ \bibnamefont
			{Stechmann}},\ }\bibfield  {title} {\bibinfo {title} {Cloud regimes as phase
			transitions},\ }\href {https://doi.org/10.1002/2016GL069396} {\bibfield
		{journal} {\bibinfo  {journal} {Geophysical Research Letters}\ }\textbf
		{\bibinfo {volume} {43}},\ \bibinfo {pages} {6579–6587} (\bibinfo {year}
		{2016})}\BibitemShut {NoStop}%
	\bibitem [{\citenamefont {Perez-Moreno}\ \emph {et~al.}(2014)\citenamefont
		{Perez-Moreno}, \citenamefont {Chavarria.},\ and\ \citenamefont
		{Ruiz-Chavarría}}]{Chavarria20}%
	\BibitemOpen
	\bibfield  {author} {\bibinfo {author} {\bibfnamefont {S.~S.}\ \bibnamefont
			{Perez-Moreno}}, \bibinfo {author} {\bibfnamefont {M.~S.~R.}\ \bibnamefont
			{Chavarria.}},\ and\ \bibinfo {author} {\bibfnamefont {G.}~\bibnamefont
			{Ruiz-Chavarría}},\ }\bibfield  {title} {\bibinfo {title} {Numerical
			solution of the swift–hohenberg equation},\ }in\ \href@noop {} {\emph
		{\bibinfo {booktitle} {Experimental and Computational Fluid Mechanics.
				Environmental Science and Engineering}}},\ \bibinfo {editor} {edited by\
		\bibinfo {editor} {\bibfnamefont {J.}~\bibnamefont {Klapp}}\ and\ \bibinfo
		{editor} {\bibfnamefont {A.}~\bibnamefont {Medina}}}\ (\bibinfo {year}
	{2014})\ pp.\ \bibinfo {pages} {409--416}\BibitemShut {NoStop}%
	\bibitem [{\citenamefont {Eguíluz}\ \emph {et~al.}(1999)\citenamefont
		{Eguíluz}, \citenamefont {Hernández-García},\ and\ \citenamefont
		{Piro}}]{Piro}%
	\BibitemOpen
	\bibfield  {author} {\bibinfo {author} {\bibfnamefont {V.~M.}\ \bibnamefont
			{Eguíluz}}, \bibinfo {author} {\bibfnamefont {E.}~\bibnamefont
			{Hernández-García}},\ and\ \bibinfo {author} {\bibfnamefont
			{O.}~\bibnamefont {Piro}},\ }\bibfield  {title} {\bibinfo {title} {Boundary
			effects in complex gizburg-landau equation},\ }\href
	{https://doi.org/10.1142/s0218127499001644} {\bibfield  {journal} {\bibinfo
			{journal} {International Journal of Bifurcation and Chaos}\ }\textbf
		{\bibinfo {volume} {9(11)}},\ \bibinfo {pages} {2209} (\bibinfo {year}
		{1999})}\BibitemShut {NoStop}%
\end{thebibliography}

\providecommand{\noopsort}[1]{}\providecommand{\singleletter}[1]{#1}%

\end{document}